\documentclass[final,1p,times]{elsarticle}
\usepackage{graphicx}
\usepackage{amssymb}
\usepackage{amsthm}

\newcommand{\beq}{\begin{eqnarray}}
\newcommand{\eeq}{\end{eqnarray}}

\begin{document}

\begin{frontmatter}

\title{Heterogeneous systems in $d$ dimensions: lower spectrum}
\author{Paolo Amore}
\ead{paolo.amore@gmail.com}
\address{Facultad de Ciencias, CUICBAS, Universidad de Colima,\\
Bernal D\'{i}az del Castillo 340, Colima, Colima, Mexico}

\begin{abstract}
We show that the properties of the lower part of the spectrum of the
Helmholtz equation for an heterogeneous system in a finite region in 
$d$ dimensions, where the solutions to the homogeneous problems are known,
can be systematically approximated by means of iterative methods. These methods
only require the specification of an arbitrary ansatz and necessarily converge
to the desired solution, regardless of the strength of the inhomogeneities, 
provided that the ansatz has a finite overlap with it.
Different boundary conditions at the borders of the domain can be assumed.
Applications in one and two dimensions are used to illustrate the methods.
\end{abstract}

\begin{keyword}
{Helmholtz equation; inhomogeneous drum; }
\end{keyword}

\end{frontmatter}

\section{Introduction}
\label{sec:intro}

In this paper we study the properties of heterogeneous systems in one or more dimensions
obeying the Helmholtz equation
\begin{eqnarray}
\left(-\Delta_d\right) \Psi_n(x_1,\dots, x_d) = E_n \  \Sigma(x_1,\dots, x_d) \ \Psi_n(x_1,\dots, x_d) 
\label{helmholtz_1}
\end{eqnarray}
where $\Delta_d$ is the Laplacian in a region $\Omega_d$ in $d$ dimensions, $\Sigma(x_1,\dots, x_d) > 0$ is the density
and $\Psi_n$ and $E_n$ are the eigenfunctions and eigenvalues of the equation respectively. The eigenfunctions
of the equation satisfy appropriate boundary conditions on $\partial\Omega_d$.

Heterogeneous systems are naturally present in nature, where the density of the medium has typically random or granular
properties,  but they can also be artificially created in the laboratory, generally with periodic properties. 
In particular the study of composite materials, built of alternating homogeneous layers of materials with different properties
(density, conductivity, refraction index, etc.), is nowadays gaining importance in different areas of Physics and Engineering.

Examples of these systems are numerous and a full account of the bibliography on the subject lies outside the scope of this paper but,
to give an idea, we can mention  composite membranes \cite{Masad96, Horgan99, Wang98,Ho00, Kang02}, periodic multi-layered 
acoustic waveguides \cite{Craster08}, elastic properties in layered media \cite{Parnell06,Smyshlyaev09,Capdeville07, Capdeville09,
Nemat-Nasser11,Srivastava12}, acoustic waves in strongly heterogeneous porous structures \cite{Rohan13} or 
photonic crystals \cite{Joannopoulos97}.
Particularly interesting, both from a physical and mathematical viewpoint, is the case of periodically layered composite materials, where the typical 
scale of the micro-structure (i.e. the period of the layers) is much smaller than the macroscopic scale of the object. The understanding of the 
macroscopic properties of such systems in terms of their microscopic features is the fundamental goal of the "homogenization method" (see 
ref.~\cite{Panasenko08} and references therein).

Our present analysis is  restricted to the class of problems described by Eq.~(\ref{helmholtz_1}), which is a subset of the general
problems listed above, although the basic ideas of the method might be used in more general situations. 
Our goal is to devise a method which allows one to obtain arbitrarily precise approximations to the eigenfunctions
and eigenvalues corresponding to the lower part of the spectrum regardless of the specific details of the problem
under consideration, such as the geometrical properties of $\Omega_d$~\footnote{For $d \neq 2$ the method requires that the solutions
to the homogeneous problem are known.}, the boundary conditions and, even more crucially, the density.

Here we describe an iterative method which progressively builds one of the eigenfunctions of the equation
above corresponding to the lower states, starting from an almost arbitrary ansatz. 

The salient features of our approach are
\begin{itemize}
\item The method applies to $d$-dimensional regions where an orthonormal basis of the Laplacian is known, with the exception
of $d=2$ where this limitation can be circumvented using a conformal mapping which transforms the original simply connected
region into a region where a basis is known;

\item Different boundary conditions on $\partial\Omega_d$ can be enforced;

\item The convergence of the method is not limited to cases where mild inhomogeneities are present: in other words, the method
is truly non--perturbative;

\item The rate of convergence can be increased using a judicious choice of the initial ansatz, which can also include 
variational parameters;

\item If the density depends on some perturbative parameter $\lambda$, and it reduces to the homogeneous case for $\lambda \rightarrow 0$, 
 the iterative method allows one to extract the perturbative expansion for the eigenfunction and eigenvalue of the targeted state to any
 order, with an appropriate number of iterations.
\end{itemize}

All these aspects are illustrated in the paper considering examples in one and two dimensions. In particular, we examine the case
of a one dimensional problem with rapidly oscillating density and different boundary conditions (we consider the cases  of
Dirichlet, mixed Dirichlet-Neumann, Neumann and periodic bc) and we obtain the {\sl exact} asymptotic behavior of the lower eigenvalues
of Eq.~(\ref{helmholtz_1}) when the size of the micro-structure goes to zero. For the case of Dirichlet bc the result contains
results previously published as a special case ~\cite{Zuazua00,Amore11}.

The paper is organized as follows:  in Section \ref{sec:method} we describe the method and prove theorems which apply 
both to the case where the spectrum of the problem does not contain a zero mode and to the case where a zero mode is present;
in Section \ref{sec:appl} we consider few applications of the method in one and two dimensions, in Section \ref{conclusions} we
draw our conclusion and discussion possible extensions of the present work.


\section{The method}
\label{sec:method}

In this section we describe different approaches which can be used to obtain 
approximations to the normal modes of the Helmholtz equation for a $d$ dimensional 
region with a variable density. The techniques that we will discuss do not require 
that the inhomogeneities are small and the convergence towards the targeted state is 
always granted. The choice of a suitable ansatz containing variational parameters 
may increase the convergence.

We first discuss the calculation of the fundamental mode and state
the following theorem:

\newtheorem{thm}{Theorem}
\begin{thm}(Spectrum without zero mode)
\label{theo1}
Consider a $d$ dimensional region of space, $\Omega_d$, and a density $\Sigma(x_1,\dots, x_d)>0$ 
which varies over $\Omega_d$; let $\Xi^{(0)}(x_1,\dots, x_d)$ be an arbitrary function defined 
on $\Omega_d$, which has an overlap with the lowest mode of the Helmholtz equation (\ref{helmholtz_1})
where $n$ is the set of numbers which specify an eigenfunction.

We assume that the spectrum of Eq.~(\ref{helmholtz_1}) does not contain a zero mode (i.e. a mode with
vanishing eigenvalue), and therefore that its eigenfunctions obey specific boundary conditions on 
$\partial\Omega_d$ (in one dimension, for instance, only Dirichlet boundary conditions at both ends, 
or mixed Dirichlet and Neumann boundary conditions at each end are allowed).

Let $G(x_1,\dots, x_d; y_1,\dots, y_d)$ be the Green's function of the homogeneous problem on $\Omega_d$ 
obeying the same boundary conditions of $\Psi$. We call $\Psi_0(x_1,\dots, x_d)$ the
eigenfunction of the fundamental mode.

Then, for $p\rightarrow \infty$, the sequence of functions
\beq
\Xi^{(p)}(x_1,\dots, x_d) &=& \sqrt{\Sigma(x_1,\dots, x_d)} \nonumber \\
&\cdot& \int_{\Omega_d} G(x_1,\dots, x_d; y_1,\dots, y_d)
\sqrt{\Sigma(y_1,\dots, y_d)} \ \Xi^{(p-1)}(y_1,\dots, y_d) \ d^dy
\label{sequence}
\eeq
converges to the lowest mode of Eq.(\ref{helmholtz_1}) having an overlap with $\Xi^{(0)}$; in particular, if
 $\Xi^{(0)}$ has an overlap with the fundamental mode of Eq.~(\ref{helmholtz_1}) one has
\beq
\Psi_0(x_1,\dots,x_d) = \lim_{p \rightarrow \infty} \frac{\Xi^{(p)}(x_1,\dots, x_d)}{\sqrt{\Sigma(x_1,\dots, x_d)}}
\eeq
\end{thm}

\newproof{pf}{Proof}
\begin{pf}
The proof is simple: we may cast Eq.(\ref{helmholtz_1}) in the form
\beq
\frac{1}{\sqrt{ \Sigma(x_1,\dots, x_d)}} \ \left(-\Delta_d\right) \ \frac{1}{\sqrt{ \Sigma(x_1,\dots, x_d)}}  \
\Phi_n(x_1,\dots, x_d) =  E_n \ \Phi_n(x_1,\dots, x_d)
\label{helmholtz_2}
\eeq
where $\Psi_n(x_1,\dots, x_d) = \Phi_n(x_1,\dots, x_d)/\sqrt{ \Sigma(x_1,\dots, x_d)}$.
Notice that Eq.(\ref{helmholtz_1}) and (\ref{helmholtz_2}) are isospectral and call $\hat{O}$
the operator appearing on the left hand side of the equation
\beq
\hat{O} \equiv \frac{1}{\sqrt{ \Sigma(x_1,\dots, x_d)}} \ \left(-\Delta_d\right) \ \frac{1}{\sqrt{ \Sigma(x_1,\dots, x_d)}}  
\nonumber 
\eeq

The spectrum of $\hat{O}$ is bounded from below and contains only positive eigenvalues;
the ansatz $\Xi^{(0)}(x_1,\dots , x_d)$ can be decomposed in the basis of the eigenfunctions
of Eq.(\ref{helmholtz_2})
\beq
\Xi^{(0)}(x_1,\dots , x_d) = \sum_{n} c_n \Phi_n(x_1,\dots,x_d) 
\eeq

Now notice that Eq.(\ref{sequence}) may be written as
\beq
\Xi^{(p)} &=& \hat{O}^{-1} \ \Xi^{(p-1)}
\label{sequence_2}
\eeq
where 
\beq
\hat{O}^{-1} \equiv \sqrt{ \Sigma(x_1,\dots, x_d)} \ \left(-\Delta_d\right)^{-1} \ \sqrt{ \Sigma(x_1,\dots, x_d)}  
\label{Oinv}
\eeq
is the inverse operator ($\hat{O} \hat{O}^{-1} = \hat{O}^{-1} \hat{O} = \textbf{1}$).  

The spectrum of $\hat{O}^{-1}$ is positive and bounded both from below and from above: therefore after $p$ 
repeated applications of Eq.(\ref{sequence_2}) one obtains 
\beq
\Xi^{(p)}(x_1,\dots , x_d) = \sum_{n} \frac{c_n}{E_n^p} \Phi_n(x_1,\dots,x_d) 
\eeq

In the limit $p \rightarrow \infty$ only the term with lowest eigenvalue of $\hat{O}$ with non--vanishing coefficient 
$c_n$ survives; this proves the theorem.
\end{pf}


We need to discuss separately the case of a spectrum containing a zero mode (in one dimension this case corresponds to
Neumann or periodic boundary conditions). 

\begin{thm}(Spectrum containing a zero mode)
\label{theo1zero}
We assume that the spectrum of Eq.~(\ref{helmholtz_1}) contains a zero mode.
Consider a $d$ dimensional region of space, $\Omega_d$, and a density $\Sigma(x_1,\dots, x_d)>0$ 
which varies over $\Omega_d$ and let $\Xi^{(0)}(x_1,\dots, x_d)$ be an arbitrary function defined 
on $\Omega_d$, which has an overlap with the lowest mode of the Helmholtz equation
(\ref{helmholtz_1}) with positive eigenvalue.

Let $G^{(0)}(x_1,\dots, x_d; y_1,\dots, y_d)$ be the "regularized" Green's 
function of the homogeneous problem on $\Omega_d$, obeying the same boundary conditions of $\Psi$. 
We call $\Psi_0(x_1,\dots, x_d)$ the eigenfunction of the zero mode, corresponding to the vanishing eigenvalue, 
and  $\Psi_1(x_1,\dots, x_d)$ the eigenfunction of the mode with smallest positive eigenvalue.

Then, for $p\rightarrow \infty$, the sequence of functions
\beq
\tilde{\Xi}^{(p)}(x_1,\dots, x_d) &=& \sqrt{\Sigma(x_1,\dots, x_d)} \nonumber \\
&\cdot& \int_{\Omega_d} G^{(0)}(x_1,\dots, x_d; y_1,\dots, y_d)
\sqrt{\Sigma(y_1,\dots, y_d)} \ \Xi^{(p-1)}(y_1,\dots, y_d) \ d^dy  \label{sequence2a} \\
\Xi^{(p)}(x_1,\dots, x_d) &=& \tilde{\Xi}^{(p)}(x_1,\dots, x_d) \nonumber \\
&-& \sqrt{\Sigma(x_1,\dots, x_d)} \ \frac{\int_{\Omega_d} \sqrt{\Sigma(y_1,\dots, y_d)} \ 
\tilde{\Xi}^{(p)}(y_1,\dots, y_d) \ d^dy}{\int_{\Omega_d}
\Sigma(y_1,\dots, y_d)  \ d^dy}
\label{sequence2b}
\eeq
converges to the lowest mode of Eq.(\ref{helmholtz_1}) with positive eigenvalue having an overlap with $\Xi^{(0)}$; in particular, if
 $\Xi^{(0)}$ has an overlap with the  the mode with smallest positive eigenvalue one has
\beq
\Psi_1(x_1,\dots,x_d) = \lim_{p \rightarrow \infty} \frac{\Xi^{(p)}(x_1,\dots, x_d)}{\sqrt{\Sigma(x_1,\dots, x_d)}}
\eeq
\end{thm}

\begin{pf}

We need to generalize the previous proof. First we notice that the explicit form of the eigenfunction of the
zero mode is known:
\beq
\Psi_0(x_1,\dots,x_d) = 1/\sqrt{V_\Omega}
\label{FM1}
\eeq
where $V_\Omega$ is the volume of $\Omega_d$. This implies that
\beq
\Phi_0(x_1,\dots,x_d) =  \sqrt{\frac{\Sigma(x_1,\dots,x_d)}{V_\Omega}} 
\label{FM2}
\eeq

As we have discussed in Ref.~\cite{Amore14}, it is convenient
to define the modified operator
\beq
\hat{O}_\gamma \equiv \frac{1}{\sqrt{\Sigma(x_1,\dots,x_d)}} (-\Delta + \gamma) \frac{1}{\sqrt{\Sigma(x_1,\dots,x_d)}}
\eeq
where $\gamma \rightarrow 0^+$; similarly we modify Eq.~(\ref{helmholtz_1}) as
\beq
(-\Delta + \gamma) \Psi_n(x_1,\dots, x_d) = E_n \Sigma(x_1,\dots, x_d) \Psi_n(x_1,\dots, x_d) 
\label{Helmholtz1b}
\eeq
The presence of a zero mode is now avoided, as long as  $\gamma >0$.

The Green's function is now
\beq
G_\gamma(x_1,\dots,x_d,y_1,\dots,y_d ) = \frac{1}{V_\Omega \gamma} + \sum^\prime_n \frac{\phi_n(x_1,\dots,x_d) \phi_n(y_1,\dots,y_d)}{\epsilon_n+\gamma}
\eeq
where  $V_\Omega$ is the volume of the region $\Omega$. 
Here $\epsilon_n$ and $\phi_n$ are the eigenvalues and eigenfunctions of the negative Laplacian on $\Omega$ 
obeying specific boundary conditions and $\sum_n^\prime$ is the sum over all possible modes, with the exclusion
of the zero mode.

In particular, in one dimension the Green's function can be expressed in a simple form which reads
\begin{eqnarray}
G_\gamma(x,y) =
\left\{ 
\begin{array}{ccc}
\frac{\left(e^{\sqrt{\gamma } (a+2 x)}-1\right) \left(e^{a
   \sqrt{\gamma }}-e^{2 \sqrt{\gamma } y}\right) \
   {\rm csch}\left(a \sqrt{\gamma }\right)
   e^{-\sqrt{\gamma } (a+x+y)}}{4 \sqrt{\gamma }} & , & -a/2 \leq x<y \leq a/2  \\
\frac{\left(e^{a \sqrt{\gamma }}-e^{2 \sqrt{\gamma }
   x}\right) \left(e^{\sqrt{\gamma } (a+2 y)}-1\right) \
   {\rm csch} \left(a \sqrt{\gamma }\right)
   e^{-\sqrt{\gamma } (a+x+y)}}{4 \sqrt{\gamma }} & , & -a/2 \leq y<x \leq a/2  \\
\end{array}
\right.
\label{GFgamma2}
\end{eqnarray}

It is possible to expand the Green's function around $\gamma=0$ as
\beq
G_\gamma(x_1,\dots,x_d,y_1,\dots,y_d ) = \frac{1}{V_\Omega \gamma} + 
\sum_{q=0}^\infty  (-1)^q \gamma^q G^{(q)}(x,y)
\eeq
where
\beq
G^{(q)}(x,y) \equiv \sum_{n\neq 0} \frac{\phi_n(x_1,\dots,x_d) \phi_n(y_1,\dots,y_d)}{\epsilon_n^{q+1}} 
\label{gq}
\eeq
Observe that $G^{(0)}(x,y)$ is the "regularized" Green's function discussed in \cite{Amore13a,Amore13b}.

At this point, we have converted our problem to the same form of the case discussed before;
the application of Theorem \ref{theo1} using the shifted Green's function $G_\gamma(x_1,\dots,x_d,y_1,\dots,y_d )$ 
could therefore be used to obtain the power series expansion of the eigenvalue and of the  eigenfunction of the 
fundamental mode around  $\gamma = 0$. However, while this calculation is useful in the evaluation of the 
sum rules involving eigenvalues of $\hat{O}$ (see the discussion in \cite{Amore14}), in the present case 
we are interested in calculating the first excited state of $\hat{O}$.

Let us first apply Theorem \ref{theo1} to the shifted problem writing
\beq
\tilde{\Xi}^{(1)}(x_1,\dots, x_d) &=& \sqrt{\Sigma(x_1,\dots, x_d)} \nonumber \\
&\cdot& \int_{\Omega_d} G_\gamma(x_1,\dots, x_d; y_1,\dots, y_d)
\sqrt{\Sigma(y_1,\dots, y_d)} \ \Xi^{(0)}(y_1,\dots, y_d) \ d^dy  
\label{sequence0}
\eeq

Assuming that the starting ansatz is independent of $\gamma$,
one should realize that that $\tilde{\Xi}^{(1)}(x_1,\dots, x_d)$ now
contains terms which depend on $1/\gamma$. These terms, however, are only
present if $\tilde{\Xi}^{(0)}(x_1,\dots, x_d)$ is not orthogonal 
to $\Phi_0(x_1,\dots,x_d)$. The effect of orthogonalizing 
$\tilde{\Xi}^{(1)}(x_1,\dots, x_d)$ with respect to 
$\Phi_0(x_1,\dots,x_d)$ is then to eliminate the factors which
depend on $1/\gamma$. Moreover, if these terms are not present, 
one can then work with the regularized Green's function $G^{(0)}$ rather
than the full Green's function: as a matter of fact the only way that
the infinitesimal contributions of $G^{(q)}$, with $q=1,2,\dots$, could survive for $\gamma\rightarrow 0^+$ would 
be if they could interfere with divergent contributions going as $1/\gamma^{q}$, which however
are absent because of the orthogonalization.

As a result we can write Eq.(\ref{sequence0}) in terms of the regularized Green's function
$G^{(0)}$ and Eqs.(\ref{sequence2a}) and (\ref{sequence2b}) follow.

At each iteration the lowest energy components of the ansatz are inflated, and the component corresponding to
the zero mode is removed by orthogonalization. Therefore, we have
\beq
\Psi_1(x_1,\dots,x_d) = \lim_{p \rightarrow \infty} \frac{\Xi^{(p)}(x_1,\dots, x_d)}{\sqrt{\Sigma(x_1,\dots, x_d)}}
\eeq

This proves the theorem.

\end{pf}


\newdefinition{rmk}{Remark}
\begin{rmk}
\label{rmk3}
Theorems \ref{theo1} and  \ref{theo1zero}  are the extension of the {\sl Power Method} (PM) for approximating
eigenvalues and eigenvectors of a finite hermitian matrix to the case of a hermitian
operator (and therefore to the infinite dimensional case).
This observation was already made in refs.\cite{Amore10,Amore10b} for the analogous results
discussed there.

As for the case of the Power Method, the rate of convergence of the iterated function
towards the eigenfunction of the fundamental mode of Eq.(\ref{helmholtz_1}) is 
essentially determined by the separation between the two largest eigenvalues of $\hat{O}^{-1}$.

The results contained in  Refs.~\cite{Amore10b} and \cite{Amore10} were limited to 
Dirichlet boundary conditions and were not formulated explicitly in terms of Green's functions.
It is convenient to consider this alternative approach in more detail: as we have discussed in Ref.~\cite{Amore12}
it is possible to cast eq.(\ref{helmholtz_1}) into the alternative form 
\beq
\left(-\Delta_d\right)^{1/2} \frac{1}{\Sigma(x_1,\dots, x_d)} \ \left(-\Delta_d\right)^{1/2} \ 
\Xi_n(x_1,\dots, x_d) =  E_n \ \Xi_n(x_1,\dots, x_d)
\label{helmholtz_2b}
\eeq
where $\Xi_n(x_1,\dots,x_d) \equiv \left(-\Delta_d\right)^{1/2} \Psi_n(x_1,\dots,x_d)$. Notice that
Eqs.(\ref{helmholtz_1}), (\ref{helmholtz_2}) and (\ref{helmholtz_2b}) are isospectral.
Following the notation of Ref.~\cite{Amore12} we define the operator
\beq
\hat{W} \equiv \left(-\Delta_d\right)^{1/2} \frac{1}{\Sigma(x_1,\dots, x_d)} \ \left(-\Delta_d\right)^{1/2} 
\eeq
Clearly the inverse operator is
\beq
\hat{W}^{-1} \equiv \left(-\Delta_d\right)^{-1/2} \Sigma(x_1,\dots, x_d) \ \left(-\Delta_d\right)^{-1/2} 
\eeq

Therefore, the implementation of Theorem 1 in matrix form may be done either the matrix elements of either
operators $\hat{O}^{-1}$ or $\hat{W}^{-1}$. The obvious choice is to use a basis of eigenstates of the 
Laplacian operator (we call $ | n_1, \dots, n_d \rangle$ and $\epsilon_{n_1,\dots,n_d}$ the eigenstates and eigenvalues
of the Laplacian respectively) and therefore 
\beq
\langle n_1, \dots , n_d| \hat{O}^{-1} | m_1, \dots, m_d \rangle &=& \sum_{r_1,\dots,r_d} \frac{\langle n_1, \dots , n_d| 
\sqrt{\Sigma} | r_1, \dots, r_d \rangle \langle r_1, \dots , r_d| \sqrt{\Sigma} | m_1, \dots, m_d \rangle}{\epsilon_{r_1,\dots,r_d}} \nonumber \\
\langle n_1, \dots , n_d| \hat{W}^{-1} | m_1, \dots, m_d \rangle &=& \frac{\langle n_1, \dots , n_d| \Sigma | m_1, \dots, m_d \rangle}{\sqrt{\epsilon_{n_1,\dots,n_d}} \ 
\sqrt{\epsilon_{m_1,\dots,m_d}}} \nonumber 
\eeq

Working with a finite portion of the Hilbert space, containing the lowest $N$ states, one obtains approximate 
values for the eigenvalues and eigenstates of the operators in terms of the corresponding quantities of the 
finite matrices. This approach amounts to applying the Rayleigh-Ritz method to this problem. Although for 
$N \rightarrow \infty$ the two matrices must yield the same eigenvalues, the approximations obtained
with a finite portion of the Hilbert space are necessarily poorer in the first case 
where the matrix element contains a sum over internal quantum numbers.

The matrix form of Theorem 1 corresponds to applying the power method directly to the $N \times N$ matrices
discussed above, picking an arbitrary trial vector. This approach only requires matrix vector multiplications,
and therefore it can be efficiently implemented in a computer, with a minimal use of memory.
\end{rmk}

\label{rmk1}
\begin{rmk}
These theorems apply to arbitrary regions in $d$-dimensions, although they require the knowledge of the 
Green's function of the homogeneous problem corresponding to that region, and therefore of the spectrum of the 
associated homogeneous problem. In two dimensions, however, one can always reduce the original problem 
to an equivalent problem over a suitable region (i.e. a region where a basis of functions is known)
with  the use of an appropriate conformal map.
 
We briefly discuss this point: consider an arbitrary simply connected region of the plane, $\Gamma$, 
where the Helmholtz equation
\beq
- \left( \frac{\partial^2}{\partial u^2} +  \frac{\partial^2}{\partial v^2} \right) \ \psi_n(u,v) = E_n \ \psi_n(u,v)
\label{Helmholtz_3}
\eeq
is fulfilled. Now let $w \equiv u + i v = f(z)$ map conformally a region $\Omega$ of the plane 
where the Green's function of the homogeneous problem is known (rectangle, circle, \dots) to $\Gamma$: on $\Omega$  
we have
\beq
- \left( \frac{\partial^2}{\partial x^2} +  \frac{\partial^2}{\partial y^2} \right) \ \Psi_n(x,y) = E_n \ \Sigma(x,y) \ \Psi_n(x,y)
\label{Helmholtz_4}
\eeq 
where $\Sigma(x,y) = \left| \frac{df}{dz} \right|^2$. Therefore we may obtain the eigenfunction and eigenvalue
of the fundamental mode of Eq.(\ref{Helmholtz_3}), applying Theorems \ref{theo1} and \ref{theo1zero} to Eq.(\ref{Helmholtz_4}).

In this particular case, the Rayleigh-Ritz approach with a finite number of basis functions is essentially
equivalent to the "conformal mapping method" of Ref.~\cite{Robnik84} (see also Ref.~\cite{Amore10b}).

\end{rmk}

\begin{rmk} {\sl (Variational theorem)}
\label{rmk2}
Let $\Xi^{(p)}(x_1,\dots,x_d)$ be the function obtained using Theorem \ref{theo1} after $p$ iterations: the expectation value of $\hat{O}$ 
(Rayleigh quotient) is
\beq
\langle \hat{O} \rangle_p =  \frac{\int d^dx \ \Xi^{(p)}(x_1,\dots,x_d) \  \hat{O} \ \Xi^{(p)}(x_1,\dots,x_d)}{\int d^dx \ \left[\Xi^{(p)}(x_1,\dots,x_d)\right]^2}
\label{evariational}
\eeq

If $\Xi^{(0)}(x_1,\dots,x_d)$ depends on one or more arbitrary parameters $\mu_i$, the minimization of $\langle \hat{O} \rangle_p$ 
with respect to these parameters leads to an optimal approximation of the eigenfunction of the fundamental mode of $\hat{O}$ in the 
class of functions considered. 
\end{rmk}


We will now state another theorem which allows one to obtain approximations to the 
eigenfunction of the fundamental mode of Eq.(\ref{helmholtz_1}). We restrict to the case
in which the zero mode is not present. A generalization to the case in which the zero mode
is present can be done along the lines of what done in Theorem \ref{theo1zero}.

\begin{thm}
\label{theo1b}
We assume that the hypotheses of  Theorem \ref{theo1} hold. 
Let $\xi^{(0)}(x_1, \dots, x_d)$ be an arbitrary function on $\Omega$, that we require to be normalized,
$\int_\Omega d^dx \  \xi^{(0)}(x_1,\dots , x_d)^2  = 1$; we adopt the Dirac bra-ket notation and work 
with the state $|\xi^{(0)} \rangle$ ($\langle \xi^{(0)} | \xi^{(0)} \rangle = 1$);

We define the state
\beq
| \chi^{(0)} \rangle &\equiv& \frac{1}{\upsilon^{(0)}} \left( \hat{O}^{-1} -  \eta^{(0)} \right) | \xi^{(0)} \rangle 
\eeq
where
\beq
\eta^{(0)} &\equiv& \langle \xi^{(0)} | \hat{O}^{-1} | \xi^{(0)} \rangle \\
\upsilon^{(0)} &\equiv& \langle \xi^{(0)} | \hat{O}^{-1} | \chi^{(0)} \rangle = \sqrt{\langle \xi^{(0)} | \hat{O}^{-2} | \xi^{(0)} \rangle - \left(\eta^{(0)}\right)^2 }
\eeq

Then, for $p \rightarrow \infty$, with $p$ integer, the state
\beq
|\xi^{(p+1)} \rangle &\equiv& 
\frac{1}{\sqrt{2}} \left\{ \sqrt{\frac{-\varepsilon^{(p)}+\eta^{(p)}+\Delta^{(p)}}{\Delta^{(p)}}} \ 
| \xi^{(p)} \rangle + \sqrt{\frac{\varepsilon^{(p)}-\eta^{(p)}+\Delta^{(p)}}{\Delta^{(p)}}} \ | \chi^{(p)} \rangle
\right\} 
\label{iter}
\eeq
with
\beq
| \chi^{(p)} \rangle &\equiv& \frac{1}{\upsilon^{(p)}} \left( \hat{O}^{-1} -  \eta^{(p)} \right) | \xi^{(p)} \rangle 
\eeq
and
\beq
\eta^{(p)} &\equiv& \langle \xi^{(p)} | \hat{O}^{-1} | \xi^{(p)} \rangle \\
\upsilon^{(p)} &\equiv& \langle \xi^{(p)} | \hat{O}^{-1} | \chi^{(p)} \rangle = 
\sqrt{\langle \xi^{(p)} | \hat{O}^{-2} | \xi^{(p)} \rangle - \left(\eta^{(p)}\right)^2 } \\
\varepsilon^{(p)} &\equiv& \langle \chi^{(p)} | \hat{O}^{-1} | \chi^{(p)} \rangle \\
\Delta^{(p)} &\equiv&  \sqrt{\left(\eta^{(p)}-\varepsilon^{(p)}\right)^2 + \left(2 \upsilon^{(p)}\right)^2}
\eeq
converges to the eigenstate of Eq.~(\ref{helmholtz_2}) with lowest eigenvalue.
\end{thm}

\begin{pf}
This theorem is an implementation of the Lanczos algorithm~\cite{Lanczos50} to Hermitian operators in an Hilbert space.
In a different context, implementations of the Lanczos algorithm are found, for example, in 
Refs.~\cite{Berger77,Dagotto85,Dagotto86}. 

In particular, it is easy to adapt the demonstration in Ref.~\cite{Berger77} to the present case:
our demonstration differs in that the method is applied to the inverse operator $\hat{O}^{-1}$.

First we notice that $|\chi^{(p)}\rangle$ is orthogonal to $|\xi^{(p)}\rangle$ and normalized to $1$:
\beq
\langle \xi^{(p)} | \chi^{(p)} \rangle = 0  \ \ \ , \ \ \ \langle \chi^{(p)} | \chi^{(p)} \rangle = 1 \nonumber
\eeq

These two states define a two-dimensional subspace and one can diagonalize $\hat{O}^{-1}$ in this space obtaining two
new states
\beq
| V_1 \rangle &\equiv& \frac{1}{\sqrt{2}} \left\{ -\sqrt{\frac{\varepsilon^{(p)}-\eta^{(p)}+\Delta^{(p)}}{\Delta^{(p)}}} \ 
| \xi^{(p)} \rangle + \sqrt{\frac{-\varepsilon^{(p)}+\eta^{(p)}+\Delta^{(p)}}{\Delta^{(p)}}} \ | \chi^{(p)} \rangle
\right\} \\
| V_2 \rangle &\equiv& \frac{1}{\sqrt{2}} \left\{ \sqrt{\frac{-\varepsilon^{(p)}+\eta^{(p)}+\Delta^{(p)}}{\Delta^{(p)}}} \ 
| \xi^{(p)} \rangle + \sqrt{\frac{\varepsilon^{(p)}-\eta^{(p)}+\Delta^{(p)}}{\Delta^{(p)}}} \ | \chi^{(p)} \rangle
\right\} 
\eeq
with corresponding eigenvalues
\beq
\mathcal{E}_1 &\equiv&  \frac{1}{2} \left( \eta^{(p)} + \varepsilon^{(p)} - \Delta^{(p)}\right) \\
\mathcal{E}_2 &\equiv&  \frac{1}{2} \left( \eta^{(p)} + \varepsilon^{(p)} + \Delta^{(p)}\right) 
\eeq

Now notice that $\Delta^{(p)} \geq | \varepsilon^{(p)} - \eta^{(p)} |$ and one finds that
\beq
\mathcal{E}_1  \leq \eta^{(p)} \ \ \ , \ \ \  \mathcal{E}_2  \geq \eta^{(p)}  \nonumber
\eeq

Therefore, Eq.(\ref{iter}) corresponds to selecting the second eigenvector, where the expectation value of
$\hat{O}^{-1}$ has increased~\footnote{Notice the difference with the standard approach, which works on the 
operator $\hat{O}$ and {\sl minimizes} its expectation value at each iteration.}. 
If this process is carried out indefinitely, then $| \xi^{(p+1)} \rangle$ must converge to the lowest state of $\hat{O}$
which has a non-zero overlap with the initial ansatz. 
\end{pf}

Also in this case one can easily apply the Theorem in its matrix form, using the matrix elements of
$\hat{O}^{-1}$ and $\hat{W}^{-1}$ that we have previously obtained.

\begin{rmk}
\label{rmk3b}
The standard implementation of the Lanczos method would work directly with the operator $\hat{O}$.  
In our case, however, there are good reasons for using $\hat{O}^{-1}$ in generating the  $|\chi^{(p)}\rangle$:
as a matter of fact, the direct application of $\hat{O}$ to an arbitrary ansatz in general 
{\sl does not}  produce a function with the same boundary conditions and therefore the method 
is not applicable in a straightforward manner. On the other hand, the application of $\hat{O}^{-1}$ to
an arbitrary function $\psi$ {\sl produces} a new function with the correct boundary conditions 
even in the case where the boundary conditions are not enforced on $\psi$.

Observe also that since the new ansatz is obtained diagonalizing $\hat{O}^{-1}$ in the 
subspace spanned by $|\xi^{(p)}\rangle$ and $|\chi^{(p)}\rangle$, one has to select the eigenstate 
with largest eigenvalue, in contrast with the case of Refs.\cite{Berger77,Dagotto85,Dagotto86}, where
the lowest eigenvalue is picked \footnote{Alternatively, one could modify the present implementation 
diagonalizing $\hat{O}$ in the subspace spanned by $|\xi^{(p)}\rangle$ and
$|\chi^{(p)}\rangle$ (in this case one would pick the eigenstate with lowest eigenvalue).}.
\end{rmk}

We now discuss the calculation of the excited states of the Helmholtz equation (\ref{helmholtz_1}) 
and prove two different theorems. 

\begin{thm}
\label{theo2} (Spectrum without a zero mode)
Let $\left\{ \Xi_0^{(1)}(x_1,\dots, x_d) , \dots , \Xi_0^{(N)}(x_1,\dots, x_d) \right\}$
be a set of arbitrary functions defined on $\Omega_d$. We assume that these functions have 
a nonzero overlap with the lowest $N$ eigenfunctions of the Helmholtz equation (\ref{helmholtz_2}) and 
that these are ordered  such that
\beq
0 &<& \frac{\int_{\Omega_d} \Xi_0^{(1)}(x_1,\dots, x_d) \ \hat{O} \ 
\Xi_0^{(1)}(x_1,\dots, x_d)  d^dx}{\int_{\Omega_d} \left[\Xi_0^{(1)}(x_1,\dots, x_d)\right]^2  d^dx} < \dots < \nonumber \\
&<&  \frac{\int_{\Omega_d} \Xi_0^{(N)}(x_1,\dots, x_d) \ \hat{O} \ \Xi_0^{(N)}(x_1,\dots, x_d)  d^dx}{\int_{\Omega_d} 
\left[\Xi_0^{(N)}(x_1,\dots, x_d)\right]^2  d^dx} 
\eeq

Consider now the new set of functions given by
\beq
\Xi_p^{(j)}(x_1,\dots, x_d) &=& \sqrt{\Sigma(x_1,\dots, x_d)} \nonumber \\
&\cdot& \int_{\Omega_d} G(x_1,\dots, x_d; y_1,\dots, y_d)
\sqrt{\Sigma(y_1,\dots, y_d)} \ \Xi^{(j)}_{p-1}(y_1,\dots, y_d)  d^dy
\label{sequence_3}
\eeq
with $j=1,2, \dots, N$ and define
\beq
\tilde{\Xi}_p^{(j)}(x_1,\dots, x_d) &=& \Xi_p^{(j)}(x_1,\dots, x_d)  \nonumber \\
&-&  \tilde{\Xi}_p^{(k)}(x_1,\dots, x_d) \ 
\sum_{k=1}^{j-1} \frac{\int_{\Omega_d} \tilde{\Xi}_p^{(k)}(y_1,\dots, y_d) \ \Xi_p^{(j)}(y_1,\dots, y_d)  
d^dy}{\int_{\Omega_d} \left[\Xi_{p}^{(k)}(y_1,\dots, y_d)\right]^2  d^dy}  \ .
\label{sequence_4}
\eeq

For $p \rightarrow \infty$ the set $\left\{ \tilde{\Xi}_p^{(1)}(x_1,\dots, x_d) , \dots, 
\tilde{\Xi}_p^{(N)}(x_1,\dots, x_d)  \right\}$ converges to the first lowest $N$ 
eigenfunctions of Eq.(\ref{helmholtz_2}).
\end{thm}

\newproof{pf2}{Proof}
\begin{pf2}
The proof is analogous to the proof of Theorem 2 of Ref.~\cite{Amore10}: at each iteration the 
functions generated with Eq.(\ref{sequence_3}) have their lower energy components inflated, since 
they have been obtained from the previous set applying the inverse operator. The first 
function of the set converges to the lowest eigenfunction, because of Theorem \ref{theo1}; the remaining
functions are orthogonalized at each step in Eq.(\ref{sequence_4}) and therefore as the number 
of iterations grows they must converge to the exact eigenfunctions of the excited modes.
\end{pf2}

\begin{rmk}
\label{rmk4} 
The generalization of this theorem to the case in which a  zero mode is present can be done straightforwardly,
along the lines discussed in Theorem \ref{theo1zero}. Notice however that this theorem is useful only for
the low lying states of an operator, since the approximate eigenfunction of the targeted state is obtained
via iteration and orthogonalization with respect to the lower states. 
An alternative approach would be modifying Theorem \ref{theo1b}, applying it to the operator 
$\hat{Q} \equiv \left( \hat{O}^{-1} - \frac{1}{\Lambda}\right)^2$. This is essentially the "folded spectrum method" (FSM)
of Refs.~\cite{Wang94,Wang96}. However, this implementation of the FSM would not be of practical use since the spectrum of the operator $\hat{Q}$,
in the part corresponding to the highly excited states of $\hat{O}$, will consist of finely spaced eigenvalues, with a
separation which tends to zero for arbitrarily high states. As a result the convergence of the method is expected to be
extremely slow for these states.
\end{rmk}

\section{Applications}
\label{sec:appl}

We will discuss some applications of the theorems that we have proved in this paper. 
The first two applications, in one and two dimensions, are simple, but serve to illustrate the 
methods discussed in this paper to high orders, obtaining highly precise results; the third application,
which concerns one dimensional heterogeneous systems with an highly oscillatory density and different
boundary conditions, will allow us to illustrate the effectiveness of our approach.

\subsection{A string with parabolic density}
Following Ref.~\cite{Amore10} we consider a string with the density 
\beq
\Sigma(x) = (1+\alpha x)^2 \ \ \ , \ \ \ |x| \leq a/2
\label{parabolic}
\eeq
with $\alpha \leq 2/a$. The choice of this particular density is dictated
uniquely by the need of illustrating the implementation of our theorems, 
limiting the technical difficulties. In our calculations we will assume 
$a=1$.

We first consider the case of Dirichlet boundary conditions and we choose the
ansatz
\beq
\xi^{(0)}(x) = \mathcal{N} \ \sqrt{\Sigma(x)} \ \psi_1^{(DD)}(x) 
\label{ansatzdd}
\eeq
where 
\beq
\mathcal{N} &=& \frac{2 \sqrt{3}}{\sqrt{\left(1-\frac{6}{\pi ^2}\right) \alpha ^2+12}}
\eeq
is a normalization constant and  $\psi_n^{(DD)}(x)$ are the Dirichlet eigenfunctions of the negative Laplacian 
on $|x|\leq a/2$
\beq
\psi_n^{(DD)}(x) &=& \sqrt{\frac{2}{a}} \ \sin \frac{n \pi (x+a/2)}{a} \nonumber
\eeq
The corresponding eigenvalues are
\beq
\epsilon_n^{(DD)} &=& \frac{n^2\pi^2}{a^2} \nonumber
\eeq

The Green's function in this case reads~\cite{Amore13a} 
\beq
G^{(DD)}(x,y) &=& 
\frac{(a-2 \max (x,y)) (a+2 \min (x,y))}{4 a} \nonumber
\eeq

After one iteration of Theorem \ref{theo1} we obtain
\beq
\xi^{(1)}(x) &=& \frac{\sqrt{2}}{\pi ^4} \mathcal{N} \ \sqrt{\Sigma(x)} \ 
\left(-6 \alpha ^2 \cos (\pi  x)+2 \pi  \alpha  (-2 (\alpha  x+1) \sin (\pi  x)+\alpha +4 x) \right. \nonumber \\
&+& \left. (\pi  \alpha  x+\pi)^2 \cos (\pi  x)\right)
\eeq

It is straightforward to carry out the iterations to larger orders, although the explicit form of the 
functions $\xi^{(p)}(x)$ becomes lengthy and therefore we do not report them here. The energy of the fundamental mode
may be approximated using the variational estimate of Eq.~(\ref{evariational}).

The expressions obtained using the variational formula take the form of a Pad\'e approximant in 
terms of the parameter $\alpha$ of the density; for instance
\beq
\langle\hat{O} \rangle_1 &=& \frac{\mathcal{A}_1}{\mathcal{B}_1} \nonumber\\
\mathcal{A}_1 &\equiv& \left(-57120 \pi ^4+5040 \pi ^6+84 \pi ^8\right) \alpha ^4 \nonumber \\
&+& \left(-860160 \pi ^4+60480 \pi ^6+3360 \pi ^8\right) \alpha^2+6720 \pi ^8 \nonumber \\
\mathcal{B}_1 &\equiv& \left(5528880-652680 \pi ^2+9226 \pi ^4+15 \pi ^6\right) \alpha ^6 \nonumber \\
&+& \left(82575360-11702880 \pi ^2+326032 \pi ^4+1260 \pi^6\right) \alpha ^4 \nonumber \\
&+& \left(-5160960 \pi ^2+451360 \pi ^4+8400 \pi ^6\right) \alpha ^2+6720 \pi ^6\nonumber
\eeq

We do not report the similar expressions for higher orders, which we have explicitly calculated, because of their
length.

In Table \ref{table1} we display the approximate values of the fundamental eigenvalue, calculated
to different orders with the aid of Theorem \ref{theo1}, using the ansatz (\ref{ansatzdd}). The remaining 
rows contain the values obtained applying repeated Shanks transformations to these values. In particular, 
notice that the values corresponding to $\alpha=1$ agree with the value obtained in Ref.~\cite{Amore10}, 
applying Theorem 1 of that paper (which is a special case of our Theorem \ref{theo1}) to order 20.

\begin{table}[tbp]
\caption{Energy of the fundamental mode of the string with density (\ref{parabolic}) for
different values of $\alpha$ using Theorem \ref{theo1}. The terms $s^{(n)}_j$ are the values
obtained after $n$ repeated Shanks transformations.}
\bigskip
\label{table1}
\begin{center}
\begin{tabular}{|c|c|c|c|}
\hline
 & $\alpha=1/2$ & $\alpha=1$ & $\alpha=2$ \\
\hline
$\langle\hat{O} \rangle_1$  &  9.69310365089956 &  9.21037410544234 &  7.76924315857119\\ 
$\langle\hat{O} \rangle_2$   &  9.68737359122776 &  9.19238760347104 &  7.73502742410347 \\
$\langle\hat{O} \rangle_3$   &  9.68702664891476 &  9.19138111461443 &  7.73341903683425 \\
$\langle\hat{O} \rangle_4$   &  9.68700556297997 &  9.19132401578685 &  7.73334058319427 \\
$\langle\hat{O} \rangle_5$   &  9.68700428051026 &  9.19132076814843 &  7.73333673246089 \\
$\langle\hat{O} \rangle_6$   &  9.68700420249783 &  9.19132058333956 &  7.73333654324603 \\
$\langle\hat{O} \rangle_7$   &  9.68700419775222 &  9.19132057282190 &  7.73333653394665 \\
$\langle\hat{O} \rangle_8$   &  9.68700419746353 &  9.19132057222331 &  7.73333653348959 \\
$\langle\hat{O} \rangle_9$   &  9.68700419744597 &  9.19132057218925 &  7.73333653346713 \\
$\langle\hat{O} \rangle_{10}$  &  9.68700419744490 &  9.19132057218731 &  7.73333653346602 \\
\hline
$s^{(1)}_1$  &  9.68700428845695 &  9.19132145507167 &  7.73333970165889 \\
$s^{(1)}_2$  &  9.6870041985241 &  9.19132058171233 &  7.73333656016117 \\
$s^{(1)}_3$  &  9.6870041974577 &  9.191320572291 &  7.7333365336999 \\
$s^{(1)}_4$  &  9.68700419744499 &  9.19132057218833 &  7.73333653346805 \\
$s^{(1)}_5$  &  9.68700419744484 &  9.1913205721872 &  7.73333653346599 \\
$s^{(1)}_6$  &  9.68700419744483 &  9.19132057218719 &  7.73333653346597 \\
$s^{(1)}_7$  &  9.68700419744483 &  9.19132057218719 &  7.73333653346597\\
$s^{(1)}_8$ &  9.68700419744483 &  9.19132057218719 &  7.73333653346597\\
\hline
$s^{(2)}_1$  &  9.68700419744490 &  9.19132057218826 &  7.73333653347512 \\
$s^{(2)}_2$  &  9.68700419744483 &  9.19132057218720 &  7.73333653346600 \\
$s^{(2)}_3$  &  9.68700419744483 &  9.19132057218719 &  7.73333653346597 \\
$s^{(2)}_4$  &  9.68700419744483 &  9.19132057218719 &  7.73333653346597 \\
$s^{(2)}_5$  &  9.68700419744483 &  9.19132057218719 &  7.73333653346597 \\
$s^{(2)}_6$  &  9.68700419744483 &  9.19132057218719 &  7.73333653346597 \\
\hline
$s^{(3)}_1$   &   9.68700419744483  &   9.19132057218719  &   7.73333653346597\\
$s^{(3)}_2$   &   9.68700419744483  &   9.19132057218719  &   7.73333653346597\\
$s^{(3)}_3$   &   9.68700419744483  &   9.19132057218719  &   7.73333653346597\\
$s^{(3)}_4$   &   9.68700419744483  &   9.19132057218719  &   7.73333653346597\\
\hline
\end{tabular}
\end{center}
\bigskip\bigskip
\end{table}

We have also applied Theorem \ref{theo1b} to the same ansatz and after one iteration 
we have obtained an explicit approximation to the fundamental eigenvalue.
In this case we have $E_0^{\alpha=1/2} \approx 9.687015834$, $E_0^{\alpha=1} \approx 9.191446083$ and 
$E_0^{\alpha=2} \approx 7.733951650$. Notice that these values are slightly more precise than the values
obtained after two iterations of Theorem \ref{theo1} (however a fair comparison between the two must take into 
account the different computational cost of each iteration in the two cases).

As we have already mentioned, the choice of a suitable ansatz may reduce the number of iterations
needed to reach a given accuracy: in some cases, however, it may be more convenient to pick a 
less precise ansatz if its analytical form allows one to carry out the iterations explicitly to large orders.
For instance, we may choose the simple ansatz
\beq
\xi^{(0)}(x) = \frac{\sqrt{105} }{8} (2 x+1) \left(1-4 x^2\right)
\label{ansatzdd2}
\eeq
and iterate it to obtain approximations to the lowest eigenvalue of the string (\ref{parabolic}) for 
$\alpha=2$. In this case, it is straightforward to perform a very large number of iterations, because
of the polynomial form of the functions involved, and we have easily calculated $105$ iterations, 
obtaining an estimate for the energy of the fundamental mode which is accurate to about $130$ digits.
Using the same ansatz we have also applied Theorem \ref{theo1b}, with $70$ iterations. 

\begin{figure}
\begin{center}
\bigskip\bigskip\bigskip
\includegraphics[width=8cm]{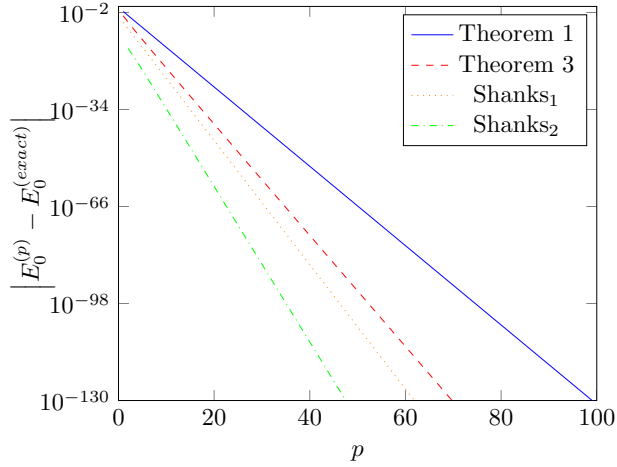}
\caption{$|E_0^{(p)}-E_0^{(exact)}|$ for the string with density $\Sigma(x) = (1+2 x)^2$ and 
Dirichlet boundary conditions obtained after $p$ iterations of Theorem \ref{theo1} (solid red curve)
and of Theorem \ref{theo1b} (dotted orange line). The initial ansatz (\ref{ansatzdd2}) is used. The remaining curves 
are obtained from one and two Shanks transformations of the first set of results.}
\label{Fig_1}
\end{center}
\end{figure}

In Fig.\ref{Fig_1} we have plotted $|E_0^{(p)}-E_0^{(exact)}|$ obtained after $p$ iterations of Theorem \ref{theo1} (solid red curve)
and of Theorem \ref{theo1b} (dotted orange line). The initial ansatz of Eq.~(\ref{ansatzdd2}) is used. The remaining curves are obtained from repeated 
Shanks transformations of the first set of results (from top to bottom the number of Shanks transformations
increases). The exact value $E_0^{(exact)}$ has been approximated with the most precise value obtained after 
$100$ iterations of  Theorem \ref{theo1}:
\beq
E_0^{(exact)} &\approx& 7.733336533465966863902638033367838303091611969871617630 \nonumber \\
                     && 20525195744620997306947223596884733603198306461387550007 \nonumber \\
                     && 5565385500030558828\dots \nonumber
\eeq

There are some important observations to make: it is clear that the rate of convergence of 
Theorem \ref{theo1b} is larger than the one of Theorem \ref{theo1}. However, the computational
cost of each iteration in the two approaches is different, since Theorem \ref{theo1b}
requires the evaluation of $\hat{O}^{-1} \xi$ and $\hat{O}^{-2} \xi$ (and the subsequent evaluation
of overlap integrals involving these functions). A second observation concerns the choice of the 
ansatz in the two approaches: in the application   of Theorem \ref{theo1} the ansatz is arbitrary 
and it does not even has to obey the boundary conditions (as a matter of fact 
$O^{-1} \xi$ obeys the boundary conditions regardless that $\xi$ does); on the other hand the ansatz
needs to obey the boundary conditions when one applies Theorem \ref{theo1b}, since the new
function is built as a linear combination of $\xi$ and $O^{-1} \xi$.
A final observation regards the use of the Shanks transformation to extract highly precise results
from a sequence of iterations: as one sees from the plot, the application of the Shanks transformation
leads to a consistent gain in convergence, which can be increased repeating  the transformation
a number of times\footnote{We have applied this technique in Ref.~\cite{Amore13a} to approximate
the fundamental eigenvalue of one dimensional systems using the exact sum rules associated to the problem.}. 
Since the results obtained from Theorem \ref{theo1} and Theorem \ref{theo1b}
are analytical and all integrals are calculated exactly, one does not have to worry about 
the insurgence of round--off errors which eventually would dominate in a purely numerical calculation.
Of course, the ability of performing a large number of iterations in the application
of our theorem depends on the form of the density, on the ansatz picked and ultimately on the dimensionality
of the problem.

We will now consider the same string, but subject to Neumann boundary conditions. We set $\alpha=2$ and
use the initial ansatz 
\beq
\xi^{(0)}(x) = (2 x+1) 
\label{ansatznn}
\eeq

\begin{figure}
\begin{center}
\bigskip\bigskip\bigskip
\includegraphics[width=8cm]{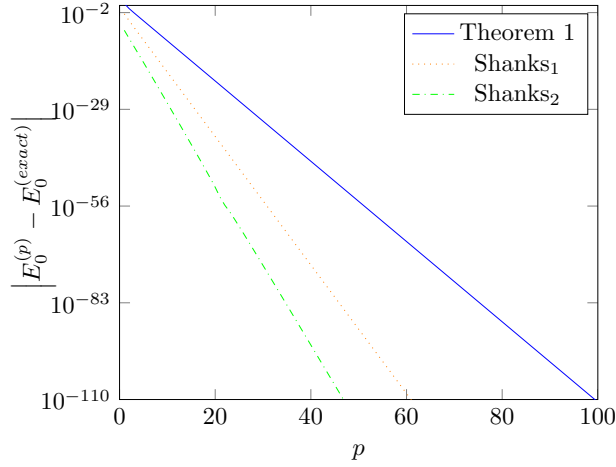}
\caption{$|E_0^{(p)}-E_0^{(exact)}|$ for the string with density $\Sigma(x) = (1+2 x)^2$ and 
Neumann boundary conditions obtained after $p$ iterations of Theorem \ref{theo1zero} (solid red curve). 
The initial ansatz (\ref{ansatznn}) is used. The remaining curves are obtained from one and two
Shanks transformations of the first set of results.}
\label{Fig_2}
\end{center}
\end{figure}

In Fig.\ref{Fig_2} we have plotted $|E_0^{(p)}-E_0^{(exact)}|$ obtained after $p$ iterations of Theorem \ref{theo1zero} (solid red curve). 
The remaining curves are obtained from repeated  Shanks transformations of the first set of results (from top to bottom the number of Shanks transformations
increases). The exact value $E_0^{(exact)}$ has been approximated with the most precise value obtained after 
$105$ iterations of  Theorem \ref{theo1zero}:
\beq
E_0^{(exact)} &\approx& 12.18713946809512900475057235600396744072004951834 \nonumber \\
                     && 58591999323057450689688696123459800390650121512407 \nonumber \\
                     && 90737876149983\dots
\eeq
where all the digits are correct.

\begin{figure}
\begin{center}
\bigskip\bigskip\bigskip
\includegraphics[width=8cm]{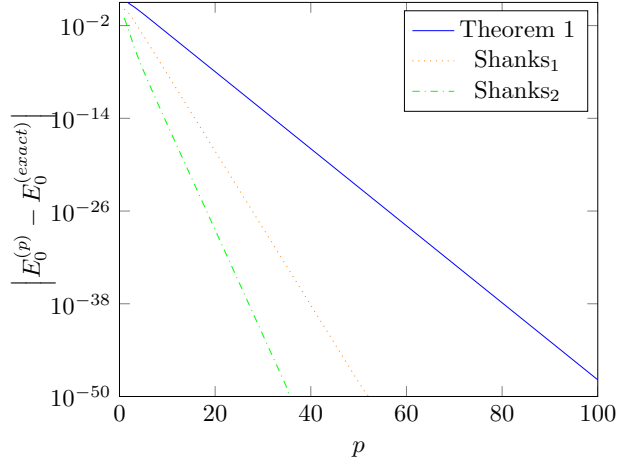}
\caption{$|E_0^{(p)}-E_0^{(exact)}|$ for the string with density $\Sigma(x) = (1+2 x)^2$ and 
periodic boundary conditions obtained after $p$ iterations of Theorem \ref{theo1zero} (solid red curve). 
The initial ansatz (\ref{ansatznn}) is used. The remainings curve are obtained from one and two
Shanks transformations of the first set of results.}
\label{Fig_3}
\end{center}
\end{figure}

We have also done a similar calculation for the case of periodic boundary conditions: we set $\alpha=2$ and use
the same initial ansatz of Eq.(\ref{ansatznn}). In Fig.\ref{Fig_3} we have plotted $|E_0^{(p)}-E_0^{(exact)}|$ 
obtained after $p$ iterations of Theorem \ref{theo1zero} (solid red curve). 
The remaining curves are obtained from repeated  Shanks transformations of the first set of results (from top to bottom the number of Shanks transformations
increases). The exact value $E_0^{(exact)}$ has been approximated with the most precise value obtained after 
$105$ iterations of  Theorem \ref{theo1zero}:
\beq
E_0^{(exact)} &\approx& 26.59255582932000052713762021489743577931546820189393\dots
\eeq
where all the digits are correct.

\subsection{A rectangular drum with density which varies along one direction}

We now apply Theorem 1 to the a two dimensional problem of a rectangular region $(x,y) \in (-a/2,a/2) \times (-b/2,b/2)$
with Dirichlet boundary conditions at the border and with density of Eq.(\ref{parabolic}).

The Green's function for this case is
\beq
G^{(D)}(x,y;x',y') &=& \sum_{n_x=1}^\infty \sum_{n_y=1}^\infty \frac{\psi^{(D)}_{n_x}(x) \phi^{(D)}_{n_y}(y) 
\psi^{(D)}_{n_x}(x') \phi^{(D)}_{n_y}(y')}{\epsilon^{(D)}_{n_x} + \eta^{(D)}_{n_y} } \ .
\label{greend1}
\eeq

Here $\psi_{n_x}(x)$ and $\phi_{n_y}(y)$ are the Dirichlet eigenfunction of the 1D negative 
Laplacian on $x \in (-a/2,a/2)$ and  $y \in (-b/2,b/2)$ respectively:
\beq
\psi_{n_x}(x) \equiv \sqrt{\frac{2}{a}} \ \sin \left( \frac{n_x\pi}{a} (x+a/2)\right) \ \ , \ \ n_x &=& 1,2,\dots \\
\phi_{n_y}(y) \equiv \sqrt{\frac{2}{b}} \ \sin \left( \frac{n_y\pi}{b} (y+b/2)\right) \ \ , \ \ n_y &=& 1,2,\dots
\eeq

$\epsilon^{(D)}_{n_x}$ and  $\eta^{(D)}_{n_y}$ are the Dirichlet eigenvalues of the 1D negative Laplacian in the
two directions:
\beq
\epsilon^{(D)}_{n_x} = \frac{n_x^2\pi^2}{a^2} \ \ \  , \ \ \ \eta^{(D)}_{n_y} = \frac{n_y^2\pi^2}{b^2} \ .
\eeq 

Eq.(\ref{greend1}) may be cast in a more convenient form as (see Refs.~\cite{Jackson98,Amore13b})
\beq
G^{(D)}(x,y;x',y') &=& \sum_{n_x=1}^\infty g_{n_x}^{(D)}(y,y') \psi^{(D)}_{n_x}(x)
\psi^{(D)}_{n_x}(x')\ ,
\label{greend2}
\eeq
where 
\beq
g_{n_x}^{(D)}(y,y') &\equiv& \sum_{n_y=1}^\infty \frac{\phi^{(D)}_{n_y}(y) \phi^{(D)}_{n_y}(y')}{\epsilon^{(D)}_{n_x} + \eta^{(D)}_{n_y} } \nonumber \\
&=& \frac{ \sinh \left( \sqrt{\epsilon_{n_x}^{(D)}} (y_{<}+b/2)\right) 
\sinh \left(\sqrt{\epsilon_{n_x}^{(D)}} (b/2-y_{>})\right) }{\sqrt{\epsilon_{n_x}^{(D)}}  
\sinh \sqrt{\epsilon_{n_x}^{(D)}} b } \ .
\label{eqgd}
\eeq
We have introduced the notation $y_< \equiv \min (y,y')$ and $y_> \equiv \max (y,y')$.

\begin{figure}
\begin{center}
\bigskip\bigskip\bigskip
\includegraphics[width=7cm]{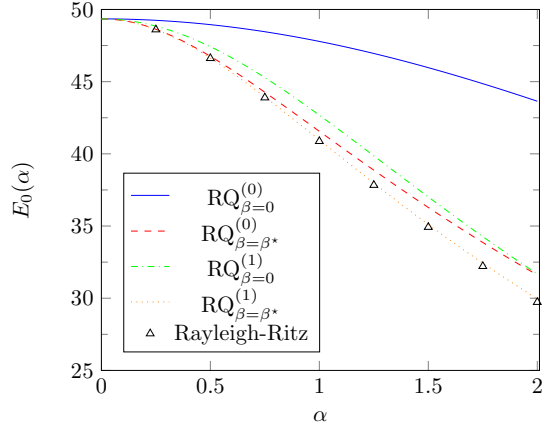}
\caption{Approximate energy of the fundamental mode of a rectangular membrane with sides $a=1$ and $b=1/2$, with 
density (\ref{parabolic}). The solid and dashed lines are the values obtained using the bounds of eqs.(\ref{bound0}) and (\ref{bound1}) respectively, setting $\beta=0$. 
The dotted and dot-dashed lines are the values obtained using the bounds of eqs.(\ref{bound0}) and (\ref{bound1}) respectively, setting $\beta=\beta^\star$.
The dots are the precise results obtained using a Rayleigh-Ritz approach with 800 basis functions.}
\label{Fig_4}
\end{center}
\end{figure}

We pick the initial ansatz
\beq
\Xi^{(0)}(x,y) = (1+\beta x) \ \sqrt{\Sigma(x,y)} \ \psi^{(D)}_{1}(x) \phi^{(D)}_{1}(y) 
\eeq

Using  this ansatz we obtain the upper bound for the energy of the fundamental mode
\beq
E_0 &\leq& E_0^{(var)}(\beta) = \frac{\int \Xi^{(0)} \hat{O} \ \Xi^{(0)}  dxdy}{\int \Xi^{(0)} \Xi^{(0)}  dxdy } \nonumber \\
&=&  \frac{20 \pi ^4 \left(6 a^2 \beta ^2 \left(b^2-a^2\right)+\pi ^2 \left(a^2+b^2\right) \left(a^2 \beta
   ^2+12\right)\right)}{a^2 b^2 \left(3 \left(120-20 \pi ^2+\pi ^4\right) a^4 \alpha ^2 \beta ^2+20 \pi
   ^2 \left(\pi ^2-6\right) a^2 \left(\alpha ^2+4 \alpha  \beta +\beta ^2\right)+240 \pi ^4\right)}
\label{bound0}
\eeq

$E_0^{(var)}(\beta)$ has a minimum at $\beta = \beta^\star$ with
\beq
\beta^\star &=&
\frac{45 a^2 \alpha ^2 \left(5 a^2+7 b^2\right)+\pi ^4 a^2 \alpha ^2 \left(a^2+b^2\right)-15 \pi ^2
   \left(2 a^4 \alpha ^2+3 b^2 \left(a^2 \alpha ^2+4\right)\right)+\sqrt{\Gamma }}{5 \left(\pi
   ^2-6\right) a^2 \alpha  \left(\left(\pi ^2-6\right) a^2+\left(6+\pi ^2\right) b^2\right)}
\eeq
where 
\beq
\Gamma &\equiv& 300 \pi ^2 \left(\pi ^2-6\right)^2 a^2 \alpha ^2 \left(a^2+b^2\right) \left(\left(\pi ^2-6\right)
   a^2+\left(6+\pi ^2\right) b^2\right) \nonumber \\
&+& \left(\left(\pi ^2-15\right)^2 a^4 \alpha ^2+\left(315-45 \pi
   ^2+\pi ^4\right) a^2 \alpha ^2 b^2-180 \pi ^2 b^2\right)^2
\eeq

Therefore $E_0^{(var)}(\beta^\star)$ provides the optimal approximation to the fundamental eigenvalue using the trial 
function $\Xi^{(0)}$.

Using Theorem 1 after one iteration we obtain the function
\beq
\Xi^{(1)}(x,y) &=&  (\alpha  x+1) \cos \left(\frac{\pi  y}{b}\right) \left\{
\frac{(a b)^{3/2} \left(\left(\pi ^2-6\right) a^2 \alpha  (\alpha +2 \beta )+12 \pi^2\right) \cos \left(\frac{\pi  x}{a}\right) }{6 \pi ^4 \left(a^2+b^2\right)} 
\right. \nonumber \\
&+& \left. (-1)^n  a^{5/2} b^{3/2}  \sum_{n=1}^\infty \left[ \frac{a \alpha (2 n+1) (\alpha +2 \beta )}{\pi ^4 n^2 (n+1)^2 \left(a^2+(2 b n+b)^2\right)} 
\cos \left(\frac{\pi  (2 n+1) x}{a}\right) \right. \right. \nonumber \\
&-& \left. \left. \frac{8 n 
   \left(\left(\pi -4 \pi  n^2\right)^2 \left(\alpha  \left(3 a^2 \alpha  \beta +8\right)+4 \beta
   \right)-48 a^2 \alpha ^2 \beta  \left(4 n^2+1\right)\right)}{\pi ^6 \left(1-4 n^2\right)^4 \left(a^2+4 b^2 n^2\right)}  \sin \left(\frac{2 \pi  n x}{a}\right) \right]\right\}
\eeq

With this function we may obtain a stricter bound for the energy of the fundamental mode of the membrane:
\beq
E_0 \leq \frac{\int \Xi^{(1)} \hat{O} \Xi^{(1)}  dxdy}{\int \Xi^{(1)} \Xi^{(1)}  dxdy } 
\label{bound1}
\eeq
where we do not report the explicit expressions for $\int \Xi^{(1)} \hat{O} \Xi^{(1)}  dxdy$ and $\int \Xi^{(1)} \Xi^{(1)}  dxdy$ because of their
lengthy form.

In Fig.\ref{Fig_1} we plot the approximate energy of the fundamental mode of a rectangular membrane with sides $a=1$ and $b=1/2$, with 
density (\ref{parabolic}). The solid and dashed lines are the values obtained using the bounds of eqs.(\ref{bound0}) and (\ref{bound1}) respectively, setting $\beta=0$. 
The dotted and dot-dashed lines are the values obtained using the bounds of eqs.(\ref{bound0}) and (\ref{bound1}) respectively, setting $\beta=\beta^\star$.
The dots are the precise results obtained using a Rayleigh-Ritz approach with 800 basis functions. This example clearly illustrates the advantages of 
using a variational approach.


\subsection{A string with rapidly oscillating density}

The iterative methods that we have described in this paper can be applied to the study of  inhomogeneous systems 
with a density  which varies on a much smaller scale  than the scale of the system itself.  
In particular we are interested in the behavior of the system when the period of the density tends to zero. 

For simplicity we restrict the present discussion to one dimension and we study the Helmholtz equation 
\beq
- \frac{d^2 \psi}{dx^2} = E_n \Sigma_{\epsilon}(x) \psi_n(x)
\label{helmholtzperiodic}
\eeq
for an inhomogeneous system with density
\beq
\Sigma_\epsilon (x) \equiv \Sigma(x/\epsilon)
\eeq
where $\Sigma(x)$ is a periodic function. The eigenfunctions $\psi_n(x)$ obey specific boundary 
conditions (Dirichlet, Neumann, periodic or mixed Dirichlet-Neumann).

The physical properties of the composite materials which can be modelled by these equations will depend
both on the microscale (i.e. the typical scale where the density changes) and on the macroscale of the system (i.e.
the typical size of the system itself). However, a numerical study of these systems is challenging, because the size 
associated to the discretization of the problem must clearly be much smaller than the size of the microscale, thus 
leading to a huge number of grid points and therefore to  an excessive use of computer memory. 
The "homogenization" method has been developed over the past forty years to obtain an effective description 
of these systems (see for example Refs.~\cite{Panasenko08, Cioranescu99, Tartar09}), which have important
applications in several areas of physics and applied mathematics.

Here we want to show that the approach described in this paper allows one to obtain systematic approximations to the 
lowest part of the spectrum of systems described by Eq.(\ref{helmholtzperiodic});
for simplicity we consider a string of unit length with a rapidly oscillating density
\beq
\Sigma_\epsilon(x) = 2  + \sin \left[\frac{2\pi}{\epsilon} (x+\eta/2)\right]
\label{density}
\eeq
where $\epsilon \rightarrow 0^+$ is the parameter determining the typical size of the oscillations of the density, and 
$\eta$ is an arbitrary phase.

This problem has been originally studied by Castro and Zuazua in Ref.~\cite{Zuazua00} for the case $\eta=1$, obtaining 
explicit expressions for the lowest eigenvalues of this string up to order $\epsilon^3$ using the WKB method: 
in Ref.~\cite{Amore11}, we have reproduced these results and we have obtained  the fundamental eigenvalue of the string 
to order $\epsilon^5$, using the method described in Ref.~\cite{Amore10}. 

\subsubsection{Dirichlet boundary conditions}

We use the same ansatz of  Ref.~\cite{Amore11}
\beq
\xi^{(0)}(x) = \sqrt{\Sigma(x)} \ \psi_1^{(DD)}(x) \ , 
\label{ansatz}
\eeq
where 
\beq
\psi_n^{(DD)}(x) \equiv \sqrt{2} \sin n\pi (x+1/2)
\eeq
are the Dirichlet eigenfunctions of the negative $1{\rm D}$ Laplacian. 

\begin{figure}
\begin{center}
\bigskip\bigskip\bigskip
\includegraphics[width=8cm]{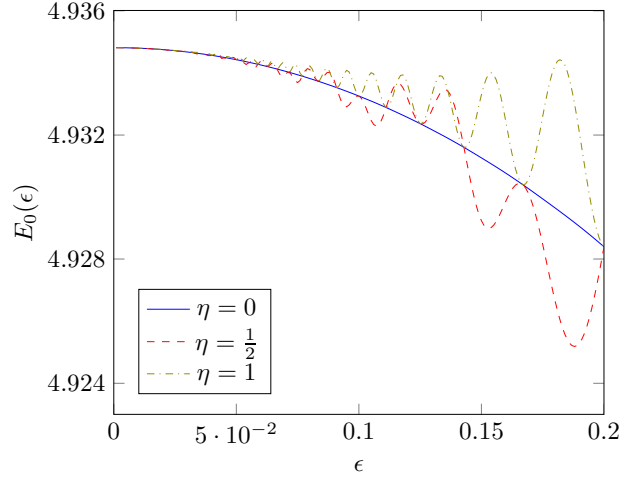}
\caption{Eq.(\ref{asym_e0}) for three different values of $\eta$.}
\label{Fig_3a}
\end{center}
\end{figure}

In analogy with the calculation in Ref.~\cite{Amore11} we apply Theorem \ref{theo1}
and obtain an expression for the lowest eigenvalue of the string which is
exact to order $\epsilon^5$, after two iterations:
\beq
E_0 &\approx&  \frac{\pi^2}{2}-\frac{1}{64} \pi ^2 \epsilon ^2 + 
\frac{1}{4} \pi  \epsilon ^3 \sin \left(\frac{\pi }{\epsilon }\right) \sin \left(\frac{\pi \eta }{\epsilon }\right)
-\frac{15 \pi ^2 \epsilon ^4}{1024} \nonumber \\
&+& \frac{1}{512} \pi  \epsilon ^5 \left(116 \sin \left(\frac{\pi }{\epsilon }\right) \sin
   \left(\frac{\pi  \eta }{\epsilon }\right)+5 \sin \left(\frac{2 \pi }{\epsilon }\right)
   \cos \left(\frac{2 \pi  \eta }{\epsilon }\right)\right) + \dots
\label{asym_e0}
\eeq
For  $\eta=1$ this result reduces to the one obtained in  Ref.~\cite{Amore11}. The behavior of $E_0$
for three different values of $\eta$ is plotted in Fig.\ref{Fig_3a}.

It is important to observe that the Rayleigh quotients obtained using Theorem \ref{theo1} form 
a monotonically decreasing sequence, which converges to the exact eigenvalue of the fundamental mode
as the number of the iterations goes to infinity. For this reason, if we study the Rayleigh quotient 
around $\epsilon=0$, we clearly have a hierarchy of contributions which correspond to the different
powers of $\epsilon$ possibly multiplied by functions which are not analytical at $\epsilon = 0$ but bounded 
(see for example eq.~(\ref{asym_e0})). At a given iteration, it may happen, depending
on the initial ansatz used, that the first few coefficients of the expansion have converged to the exact 
value: in this case, the contribution of the first term which has not yet converged must decrease at the 
next iteration. Since this must happen in a infinitesimal interval containing $\epsilon=0$, it means that 
only even terms in $\epsilon$, may decrease in the whole interval: in other words, the convergence of 
a term which is even in $\epsilon$ forces the convergence of the subsequent term in the expansion, which is odd
in $\epsilon$. This is observed in Eq.(\ref{asym_e0}) which is exact to order $\epsilon^5$.

Alternatively, one can obtain a bound on the error on the lowest eigenvalue calculating 
the mean square deviation using the function obtained after $k$ iterations, 
$\xi^{(k)}(x)$\footnote{Clearly, Theorem \ref{theo1} implies that $lim_{k \rightarrow \infty} \Delta^{(k)} = 0$.}:
\beq
\Delta^{(k)} &\equiv& \sqrt{\frac{\int \xi^{(k)}(x) \hat{O}^2 \xi^{(k)}(x) dx}{\int \left[ \xi^{(k)}(x)\right]^2 dx}
- \left[ \frac{\int \xi^{(k)}(x) \hat{O}\xi^{(k)}(x) dx}{\int \left[ \xi^{(k)}(x)\right]^2 dx}\right]^2} 
\eeq

\begin{figure}
\begin{center}
\bigskip\bigskip\bigskip
\includegraphics[width=8cm]{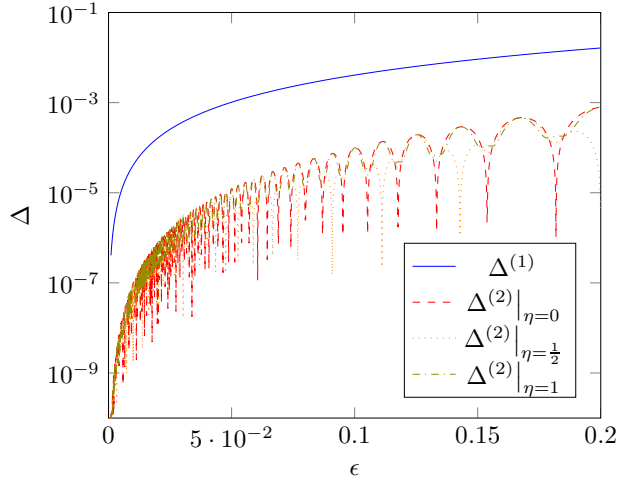}
\caption{Mean square deviations obtained with after one and two iterations for the string with density 
(\ref{density})  using the initial ansatz (\ref{ansatz}), with Dirichlet bc.}
\label{Fig_3b}
\end{center}
\end{figure}

The mean square deviations corresponding to the first two iterations are:
\beq
\Delta^{(1)} &=& \frac{1}{64} \sqrt{7} \pi ^2 \epsilon ^2 + O\left[\epsilon^3\right] \\
\Delta^{(2)} &=& \frac{\pi  \epsilon ^3}{96 \sqrt{210}} \ \left[ \left(30240-31 \pi ^6\right) \cos \left(\frac{2 \pi  \eta }{\epsilon }\right)
+ \left(16 \pi ^6-15120\right) \cos \left(\frac{2 \pi }{\epsilon }-\frac{2 \pi  \eta }{\epsilon
   }\right) \right. \nonumber \\
&+& \left. \left(16 \pi ^6-15120\right) \cos \left(\frac{2 \pi  \eta }{\epsilon
   }+\frac{2 \pi }{\epsilon }\right)+\left(30240-31 \pi ^6\right) \cos \left(\frac{2 \pi
   }{\epsilon }\right)+32 \pi ^6-30240\right]^{1/2} \nonumber \\
    &+& O\left[\epsilon^4\right] 
\eeq
These functions are plotted in Fig.\ref{Fig_3b}.

We now generalize the ansatz of eq.(\ref{ansatz}) to arbitrary states:
\beq
\xi_n^{(0)}(x) = \sqrt{\Sigma(x)} \ \psi_1^{(DD)}(x) \ .
\label{ansatz2}
\eeq

Although Theorem \ref{theo1} only holds for the lowest state, we may check if the functions obtained after one iterations
are good approximations to the excited eigenmodes for $\epsilon \rightarrow 0$;
applying Eq.(\ref{sequence}) we have
\beq
\xi_n^{(1)}(x) &=&  \sqrt{\Sigma_\epsilon(x)} \int_{-1/2}^{1/2} G^{(DD)}(x;y) \sqrt{\Sigma_\epsilon(y)} \xi_n^{(0)}(y) dy \nonumber \\ 
&=& \frac{\sqrt{\sin \left(\frac{\pi  (\eta +2 x)}{\epsilon }\right)+2}}{2 \sqrt{2} \pi ^2 n^2 \left(\epsilon ^2 n^2-4\right)^2} 
\left[ 128 \sin \left(\pi  n x+\frac{\pi  n}{2}\right) + 16 \epsilon ^2 n^2 \sin \left(\frac{1}{2} \pi  n (2 x+1)\right) \left(\sin
   \left(\frac{\pi  (\eta +2 x)}{\epsilon }\right)-4\right) \right. \nonumber \\
&+& \left. 8 \epsilon ^3 n^3 \left(-(-1)^n (2 x+1) \cos \left(\frac{\pi  (\eta +1)}{\epsilon}\right)
 +  \cos \left(\frac{\pi  (2 \epsilon  n x+\epsilon  n-2 \eta -4 x)}{2 \epsilon}\right) \right. \right. \nonumber \\
&+& \left. \left. \cos \left(\frac{\pi  (2 \epsilon  n x+\epsilon  n+2 \eta +4 x)}{2 \epsilon
   }\right)+(2 x-1) \cos \left(\frac{\pi -\pi  \eta }{\epsilon }\right)\right) \nonumber \right. \\ 
&+& \left. 4 \epsilon ^4 n^4 \sin \left(\pi  n x+\frac{\pi  n}{2}\right) \left(\sin \left(\frac{\pi
    (\eta +2 x)}{\epsilon }\right)+2\right)    \right] \nonumber 
\eeq

To see if $\xi_n^{(1)}(x)$ are good approximations we estimate the mean square deviation using these states;
if we let $\epsilon \rightarrow 0^+$ keeping $n$ fixed we have
\beq
\Delta_n^{(1)} &\approx& \frac{ \sqrt{7} }{64} \pi ^2 \epsilon ^2 n^4 + O\left(\epsilon^3\right)
\eeq

Therefore $\xi_n^{(1)}(x)$  are good approximations if $\Delta_n^{(1)} \ll 1$, implying 
$n \ll \frac{2 \sqrt{\frac{2}{\pi }}}{\sqrt[8]{7} \sqrt{\epsilon }} \approx \frac{1.25}{\sqrt{\epsilon}}$.
Working in this limit, we may estimate the eigenvalues of the lowest part of the spectrum
evaluating the expectation value of $\hat{O}$:
\beq
E_n &\approx& \langle \hat{O} \rangle_{\xi_n^{(1)}} = \frac{\int \xi_n^{(1)}(x) \hat{O}\xi_n^{(1)}(x) dx}{\int \left[ \xi_n^{(1)}(x)\right]^2 dx} \nonumber \\
&\approx& \frac{\pi ^2 n^2}{2} -\frac{1}{64} \pi ^2 \epsilon ^2 n^4 +
\frac{1}{4} \pi  \epsilon ^3 n^4 \sin \left(\frac{\pi }{\epsilon }\right) \sin
   \left(\frac{\pi  \eta }{\epsilon }\right) + O\left[\epsilon^4\right]
\eeq

For $\eta=1$ this result reproduces the estimate obtained in refs.\cite{Zuazua00,Amore11}.

\subsubsection{Neumann-Dirichlet and Dirichlet-Neumann boundary conditions}

Working in analogy to what we have done for the case of Dirichlet boundary conditions in the previous chapter, we
use the ansatz
\beq
\xi^{(0)}(x) = \sqrt{\Sigma(x)} \ \psi_1^{(ND)}(x) \ , 
\label{ansatz_nd}
\eeq
where 
\beq
\psi_n^{(ND)}(x) \equiv \sqrt{2} \sin \left(\frac{1}{4} \pi  (2 n-1) (2 x+3)\right)
\eeq
are the Dirichlet eigenfunctions of the negative $1{\rm D}$ Laplacian on $x \in (-1/2,1/2)$.

We apply Theorem \ref{theo1} and after two iterations the expressions for the eigenvalue of the fundamental mode
has converged to order $\epsilon$, for $\epsilon \rightarrow 0$:
\beq
E_0^{(ND)} \approx \frac{\pi ^2}{8}-\frac{1}{16} \pi  \epsilon  \cos \left(\frac{\pi  (1-\eta) }{\epsilon }\right) + O\left(\epsilon^2\right)
\label{asym_e0_nd}
\eeq
Notice that the convergence in this case is slower than in the case of Dirichlet boundary conditions.

Observing that under the changes $\epsilon \rightarrow -\epsilon$ and $\eta\rightarrow -\eta$, 
we have that $\Sigma(x) \rightarrow \Sigma(-x)$, we are able to obtain the energy of the lowest mode
for Dirichlet-Neumann boundary conditions for free:
\beq
E_0^{(DN)} \approx \frac{\pi^2}{8} + \frac{1}{16} \pi  \epsilon  \cos \left(\frac{\pi  (\eta +1)}{\epsilon }\right) + O\left(\epsilon^2\right)
\eeq

\begin{figure}
\begin{center}
\bigskip\bigskip\bigskip
\includegraphics[width=8cm]{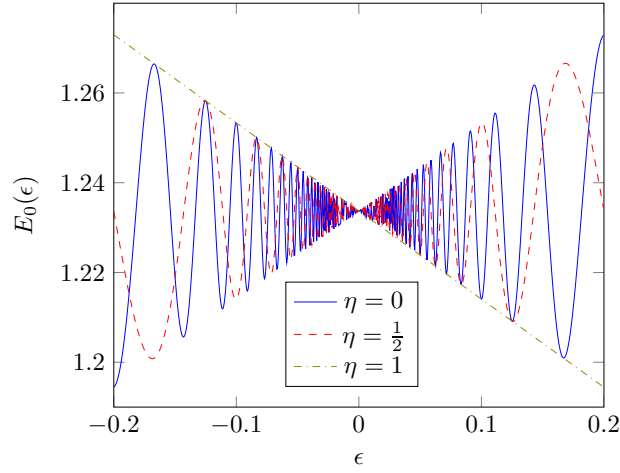}
\caption{Eq.(\ref{asym_e0_nd}) for three different values of $\eta$.}
\label{Fig_3a_nd}
\end{center}
\end{figure}

In Fig.\ref{Fig_3a_nd} we plot eq.(\ref{asym_e0_nd}) for three different values of $\eta$ (the curves for $\epsilon <0$ may be viewed 
as the asymptotic expression for $E_0^{(DN)}$ with phase $-\eta$). As for the case of Dirichlet boundary conditions we calculate 
the mean square deviations after one and two iterations, which read:
\beq
\Delta^{(1)} &=& \left| \epsilon \cos\left(\frac{\pi  (\eta -1)}{\epsilon }\right) \right| \  \frac{\pi}{64} \sqrt{\frac{1}{6} \left(\pi ^4-96\right)}  + O\left(\epsilon^2\right) \\
\Delta^{(2)} &=&    \left| \epsilon \cos\left(\frac{\pi  (\eta -1)}{\epsilon }\right) \right| \ \frac{\pi}{768}  \
\sqrt{\frac{17 \pi ^8}{70}-2304} + O\left(\epsilon^2\right)
\eeq

For a general state, we  use the initial ansatz $\xi_n^{(0)}(x) = \sqrt{\Sigma(x)} \ \psi_n^{(ND)}(x)$ and obtain the
approximate expression for the $n^{th}$ eigenvalue after one iteration
\beq
E_n^{(ND)} \approx \frac{\pi ^2 }{8} (1-2 n)^2-\frac{\pi}{16}   \epsilon  (1-2 n)^2 \cos
   \left(\frac{\pi }{\epsilon }-\frac{\pi  \eta }{\epsilon }\right)  + O\left(\epsilon^2\right)
\eeq

The corresponding mean square deviation in this case is 
\beq
\Delta_n^{(1)} = \frac{\pi   (1-2 n)^2}{64 \sqrt{6}} \sqrt{\left(\pi ^4 (1-2 n)^4-96\right)}  \left| \epsilon \cos\left(\frac{\pi  (\eta -1)}{\epsilon }\right)\right|
+  O\left(\epsilon^2\right)
\eeq
which therefore requires $n \ll \epsilon^{-1/4}$.

\subsubsection{Neumann boundary conditions}

This case is slightly more complicated since the spectrum contains a zero mode. The implementation of the iterative method 
is done following Theorem \ref{theo1zero} and using an initial ansatz orthogonal to the zero mode; we use the ansatz
\beq
\xi^{(0)}(x) &=& \sqrt{\Sigma(x)} \psi_{1,2}^{(NN)}(x) - 
\frac{\int_{-a/2}^{a/2} \psi_{1,2}^{(NN)}(x) \Sigma(x) dx }{\int_{-a/2}^{a/2} \Sigma(x) dx } \ \sqrt{\Sigma(x)} \nonumber \\
&=& \sqrt{2} \sqrt{2+\sin \left(\frac{\pi  (\eta +2 x)}{\epsilon}\right)} \ 
\left(\sin (\pi  x)-\frac{4 \epsilon  \cos\left(\frac{\pi }{\epsilon }\right) 
\cos \left(\frac{\pi\eta }{\epsilon }\right)}{\left(\epsilon ^2-4\right) 
\left(\epsilon  \sin \left(\frac{\pi }{\epsilon }\right) \sin \left(\frac{\pi  \eta }{\epsilon }\right)+2 \pi \right)}\right)
\eeq
where
\beq
\psi_{n,u}^{(NN)}(x) &=& \left\{\begin{array}{ccc}
\sqrt{\frac{1}{a}} & , & n = 0 \ , \ u=1 \\
\sqrt{\frac{2}{a}} \cos \frac{2 n \pi x}{a} & , & n>0 \ , \ u=1 \\
\sqrt{\frac{2}{a}} \sin \frac{(2 n-1) \pi x}{a} & , & n > 0 \ , \ u=2 \\
\end{array}
\right. 
\eeq
are the eigenfunction of the negative Laplacian obeying Neumann boundary conditions.

After one iteration we find
\beq
E_0^{(NN)} \approx \frac{\pi ^2}{2}-\frac{\pi  \epsilon}{2}   \sin
   \left(\frac{\pi }{\epsilon }\right) \sin \left(\frac{\pi
    \eta }{\epsilon }\right) + O\left( \epsilon^2\right)
\label{e0_nn}
\eeq

Notice that for $\eta=1$ this result agrees with the approximate expression obtained in Eq.(53) of Ref.~\cite{Amore13b}.

The mean square deviation calculated  after one iteration is
\beq
\Delta^{(1)} &\approx&  \frac{\pi |\epsilon|}{48 \sqrt{5}}  \ 
\left[ \left(-360 \cos \left(\frac{2 \pi  (\eta -1)}{\epsilon }\right)
+\left(720-7 \pi^4\right) \cos \left(\frac{2 \pi  \eta }{\epsilon}\right) \right. \right.\nonumber \\
&+& \left. \left. \cos \left(\frac{2 \pi }{\epsilon }\right)
   \left(8 \pi ^4 \cos \left(\frac{2 \pi  \eta }{\epsilon
   }\right)-7 \pi ^4+720\right)+8 \left(-45 \cos
   \left(\frac{2 \pi  (\eta +1)}{\epsilon }\right)+\pi
   ^4-90\right)\right)\right]^{1/2} \nonumber \\
  &+& O\left( \epsilon^2\right)
\eeq

\begin{figure}
\begin{center}
\bigskip\bigskip\bigskip
\includegraphics[width=8cm]{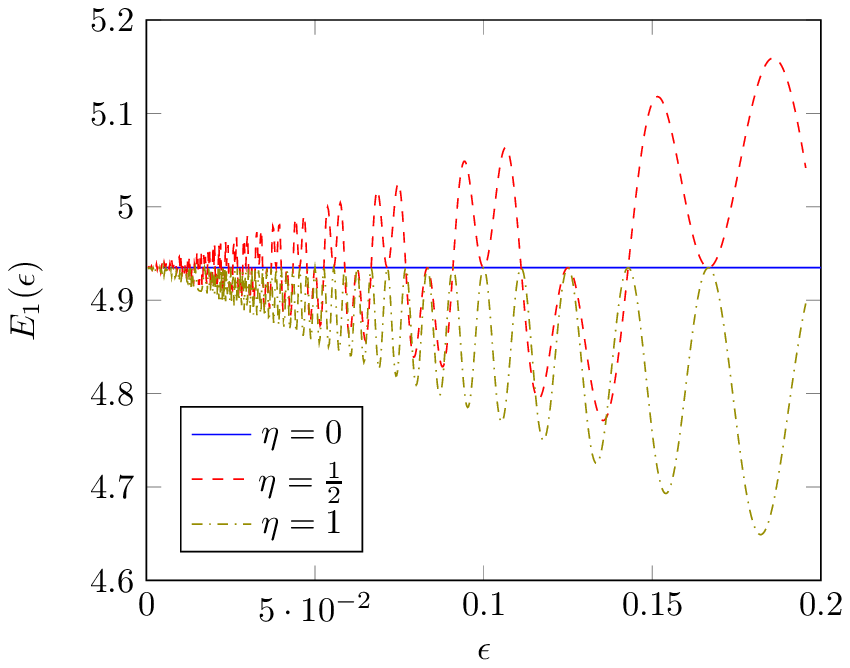}
\caption{Eq.(\ref{e0_nn}) for three different values of $\eta$.}
\label{Fig_3a_nn}
\end{center}
\end{figure}

\subsubsection{Periodic boundary conditions}

In this case we consider the ansatz
\beq
\xi^{(0)}(x) &=& \sqrt{\Sigma(x)} \left(\cos\phi \ \psi_{1,1}^{(PP)}(x)+\sin\phi \ \psi_{1,2}^{(PP)}(x) \right) \nonumber \\
&-& \frac{\int_{-a/2}^{a/2} \left(\cos\phi \ \psi_{1,1}^{(PP)}(x)+\sin\phi \ \psi_{1,2}^{(PP)}(x) \right) 
\Sigma(x) dx }{\int_{-a/2}^{a/2} \Sigma(x) dx } \ \sqrt{\Sigma(x)} \nonumber 
\eeq
where
\beq
\psi_{n,u}^{(PP)}(x) &=& \left\{\begin{array}{ccc}
\sqrt{\frac{1}{a}} & , & n = 0 \ , \ u=1 \\
\sqrt{\frac{2}{a}} \cos \frac{2 n \pi x}{a} & , & n > 0 \ , \ u=1 \\
\sqrt{\frac{2}{a}} \sin \frac{2 n \pi x}{a} & , & n > 0 \ , \ u=2 \\
\end{array}
\right.  \nonumber
\eeq
are the eigenfunctions of the 1D negative Laplacian obeying periodic boundary conditions.
Remember that these eigenfunctions are twice degenerate for $n>0$: the ansatz is expressed 
in terms of the lowest two degenerate states.

The explicit form of this ansatz is
\beq
\xi^{(0)}(x) &=& \sqrt{2} \sqrt{2+\sin \left(\frac{\pi  (\eta +2 x)}{\epsilon}\right)} \ 
\left( \sin (\phi ) \sin (2 \pi  x)+\cos (\phi ) \cos (2 \pi  x)  \right. \nonumber \\
&-& \left. \frac{\epsilon  \sin \left(\frac{\pi }{\epsilon }\right) \left(\epsilon \sin (\phi ) 
\cos \left(\frac{\pi  \eta }{\epsilon }\right)+\cos (\phi ) 
\sin \left(\frac{\pi  \eta }{\epsilon }\right)\right)}{\left(\epsilon^2-1\right) \left(\epsilon  \sin \left(\frac{\pi }{\epsilon }\right) \sin
\left(\frac{\pi  \eta }{\epsilon }\right)+2 \pi \right)}\right)
\eeq

Applying Theorem \ref{theo1zero} after one iteration, for $\epsilon \rightarrow 0$, the Rayleigh quotient provides the estimate for the 
lowest non-trivial eigenvalues
\beq
E_{1,2}^{(PP)} = 2 \pi ^2-2 \pi  \epsilon  \sin \left(\frac{\pi }{\epsilon }\right) \cos
   ^2(\phi ) \sin \left(\frac{\pi  \eta }{\epsilon }\right) + O\left(\epsilon^2\right)
\label{e0_pp}
\eeq

The mean square deviation obtained using $\xi^{(1)}(x)$, for $\epsilon \rightarrow 0$, behaves as
\beq
\Delta^{(1)} = \frac{1}{3} \sqrt{\frac{2}{5}} \pi \left| \epsilon \cos(\phi ) \sin\left(\frac{\pi}{\epsilon }\right)  \sin\left(\frac{\pi  \eta }{\epsilon }\right)\right|
 \sqrt{ \pi^4-45 \cos (2 \phi )-45} + O\left(\epsilon^2\right)
\eeq

\begin{figure}
\begin{center}
\bigskip\bigskip\bigskip
\includegraphics[width=7cm]{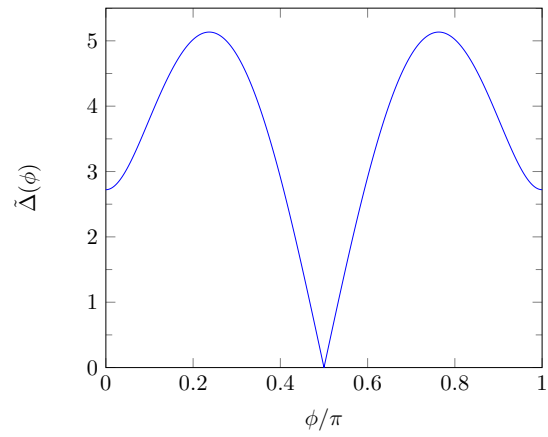}
\caption{Reduced mean square deviation $\tilde{\Delta}$ as a function of $\phi$.}
\label{Fig_9}
\end{center}
\end{figure}

\begin{figure}
\begin{center}
\bigskip\bigskip\bigskip
\includegraphics[width=8cm]{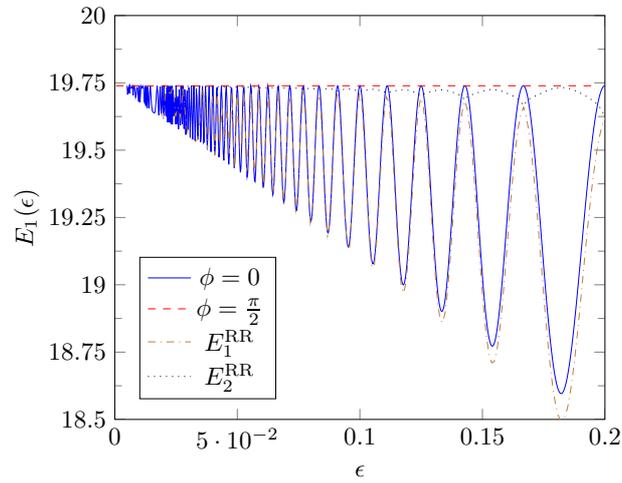}
\caption{Asymptotic behavior of $E_{1,2}(\epsilon)$ for $\eta=1$ obtained using Theorem \ref{theo1zero} and numerical results obtained with the
Rayleigh-Ritz method.}
\label{Fig_10}
\end{center}
\end{figure}

In Fig.~\ref{Fig_9}, we plot the quantity $\tilde{\Delta} \equiv  \left| \cos(\phi )\right| \sqrt{ \pi^4-45 \cos (2 \phi )-45}$ as a function
of the parameter $\phi$ which controls the mixing of the degenerate states in the initial ansatz. As we see from the figure, 
the mean square deviation goes to zero to order $\epsilon$ at $\phi=\pi/2$, while it has a local minimum at $\phi=0$.
This suggests that the two lowest degenerate states of homogeneous problem split at finite $\epsilon$ into a state with constant energy 
(to order $\epsilon$) and a state with oscillating energy. In Fig.~\ref{Fig_10} we display the asymptotic formulas (\ref{e0_pp}) 
for $\eta=1$ (corresponding to $\phi=0$ and $\phi=\pi/2$) and compare them with the numerical results obtained with the Rayleigh-Ritz method
using a set of $101$ states. The agreement between the analytical and the numerical results is evident. A similar agreement was also observed
for the cases for $\eta=0$ and $\eta=1/2$, although we do not include any figure for these cases.

\section{Conclusions}
\label{conclusions}

We have devised an iterative method which allows one to obtain arbitrarily precise solutions to the lowest eigenvalues and eigenfunctions
of the Helmholtz equation for an heterogeneous medium in $d$ dimensions, obeying different boundary conditions. The method requires the knowledge of the
Green's function for the corresponding homogeneous problem and therefore it is limited to certain class of domains (in two dimensions, however, one can apply
it to arbitrary simply connected domains, which can be conformally mapped to a circle or to a square, for which explicit expressions for the homogeneous 
Green's functions are available). A sequence of approximations which converges to the targeted solution is obtained applying iteratively the inverse operator
of Eq.~(\ref{app_A_eq2}) to an initial arbitrary ansatz, with a finite overlap with the targeted state. Some applications in one and two dimensions
are used to illustrate the method. In particular we have obtained the asymptotic behaviour of the lowest part of the spectrum of a string
with a rapidly oscillating density when the size of the micro-structure goes to zero, for Dirichlet, Dirichlet-Neumann, Neumann and periodic boundary conditions.

\appendix

\section{Inverse operator}
\label{AppA}

In this paper we have considered the operator
\beq
\hat{O} = \frac{1}{\sqrt{\Sigma({\bf x})}} (-\Delta ) \frac{1}{\sqrt{\Sigma({\bf x})}}
\label{app_A_eq1}
\eeq
We are interested in the spectrum of this operator on a $d$-dimensional region $\Omega_d$, 
where its eigenfunctions obey appropriate boundary conditions; depending on the boundary conditions
the spectrum may contain a zero mode.

Let us first consider the case in which a zero mode is not present and let $G({\bf x},{\bf y})$ be the Green's function
of the negative Laplacian on $\Omega_d$ corresponding to the specific boundary conditions. 
We assume that $G({\bf x},{\bf y})$ is known exactly (which in general is not true). 
Under these assumptions it is straightforward to check that the operator
\beq
\hat{P} f = \sqrt{\Sigma({\bf x})} \int d^dy \ G({\bf x},{\bf y}) \sqrt{\Sigma({\bf y})} f({\bf y}) 
\label{app_A_eq2}
\eeq
is the inverse of $\hat{O}$. As a matter of fact
\beq
\hat{O} \hat{P} f &=&  \frac{1}{\sqrt{\Sigma({\bf x})}} (-\Delta ) \int d^dy \ G({\bf x}, {\bf y}) \sqrt{\Sigma({\bf y})} f({\bf y}) \nonumber \\
&=&  \frac{1}{\sqrt{\Sigma({\bf x})}}  \int d^dy \  \delta({\bf x}-{\bf y}) \ \sqrt{\Sigma({\bf y})} f({\bf y}) = f({\bf x})
\eeq

The case in which a zero mode is present is more delicate. In this case we assume that $G({\bf x},{\bf y})$ is the regularized 
Green's function, which does not contain the divergent contribution of the zero mode. In this case one has
\beq
(-\Delta) G({\bf x},{\bf y}) =  \delta({\bf x}-{\bf y}) - \frac{1}{V_\Omega}
\eeq
and
\beq
\hat{O} \hat{P} f &=&  \frac{1}{\sqrt{\Sigma({\bf x})}} (-\Delta ) \int d^dy \ G({\bf x}, {\bf y}) \sqrt{\Sigma({\bf y})} f({\bf y}) \nonumber \\
&=&  \frac{1}{\sqrt{\Sigma({\bf x})}}  \int d^dy \  \left[ \delta({\bf x}-{\bf y}) - \frac{1}{V_\Omega}\right] \ \sqrt{\Sigma({\bf y})} f({\bf y}) \nonumber \\
&=& f({\bf x}) -  \frac{1}{V_\Omega} \frac{1}{\sqrt{\Sigma({\bf x})}}  \int d^dy  \ \sqrt{\Sigma({\bf y})} f({\bf y}) 
\eeq

Therefore we see that, for a generic $f$ the operator $\hat{P}$ is not the inverse of $\hat{O}$: however, if $f({\bf x})$ is orthogonal to the zero mode, 
whose eigenfunction is precisely $\sqrt{\Sigma({\bf x})}$, then it follows that $\int d^dy  \ \sqrt{\Sigma({\bf y})} f({\bf y}) =0$ and 
$\hat{P}$ behaves as $\hat{O}^{-1}$.

\section{Matrix representation}
\label{AppB}

In some situations it may be convenient to use the explicit representation of the Green's function for the homogeneous
problem. For the case where a zero mode is not present the Green's function reads
\begin{eqnarray}
G(x_1, \dots, x_d; y_1,\dots,y_d) = \sum_n \frac{\phi_n(x_1,\dots,x_d) \phi_n(y_1,\dots,y_d)}{\epsilon_n}
\end{eqnarray}
where $n$ is the set of all quantum numbers which specify an eigenstate of the homogeneous problem.

The iterations obtained applying Theorem \ref{theo1} to the initial ansatz $\Xi^{(1)} = \sqrt{\Sigma} \phi_{n_0}$ can then be obtained in the 
form of multiple series whose summands involve the matrix elements of the density:
\begin{eqnarray}
\Xi^{(1)}(x_1, \dots, x_d) &=& \sqrt{\Sigma(x_1, \dots, x_d)} \  \sum_{n_1} \frac{\langle n_1 | \Sigma | 1 \rangle}{\epsilon_{n_1}}  \phi_{n_1}(x_1,\dots,x_d) \\
\Xi^{(2)}(x_1, \dots, x_d) &=& \sqrt{\Sigma(x_1, \dots, x_d)} \  \sum_{n_1,n_2} \frac{\langle n_1 | \Sigma | n_2\rangle \langle n_2 | \Sigma | 1\rangle}{\epsilon_{n_1}\epsilon_{n_2}}  \phi_{n_1}(x_1,\dots,x_d) \\
&\dots& \nonumber \\
\Xi^{(p)}(x_1, \dots, x_d) &=& \sqrt{\Sigma(x_1, \dots, x_d)} \  \sum_{n_1,n_2, \dots, n_p} \frac{\langle n_1 | \Sigma | n_2\rangle \dots \langle n_p | \Sigma | 1\rangle}{\epsilon_{n_1} \dots \epsilon_{n_p}}  \phi_{n_1}(x_1,\dots,x_d) 
\end{eqnarray}

These expressions can be used to cast the integrals $\int_{\Omega_d} d^dx \left[ \Xi^{(p)}\right]^2$
and $\int_{\Omega_d} d^dx \left[ \Xi^{(p)} \Xi^{(p-1)}\right]$ in the form of infinite series. This implementation
of Theorem \ref{theo1} corresponds essentially to the approach of Ref.~\cite{Amore10b}.

Let us apply the operator $\hat{O} = \frac{1}{\sqrt{\Sigma}} (-\Delta) \frac{1}{\sqrt{\Sigma}}$ to the function obtained after $p$ iterations:
\beq
\hat{O} \Xi^{(p)}(x_1, \dots, x_d) &=& \frac{1}{\sqrt{\Sigma(x)}}  \sum_{n_1,n_2, \dots, n_p} \frac{\langle n_1 | \Sigma | n_2\rangle \dots \langle n_p | \Sigma | 1\rangle}{\epsilon_{n_2} \dots \epsilon_{n_p}}  \phi_{n_1}(x_1,\dots,x_d) 
\eeq
where we have used the fact that $\phi_{n_1}(x_1,\dots,x_d)$ is an eigenfunction of the negative Laplacian.
We now use the completeness of the basis of the eigenfunctions of the Laplacian to write
\beq
\sum_{n_1}  \langle n_1 | \Sigma | n_2\rangle \phi_{n_1}(x_1,\dots,x_d) = \Sigma(x)  \phi_{n_2}(x_1,\dots,x_d)
\eeq
and therefore
\beq
\hat{O} \Xi^{(p)}(x_1, \dots, x_d) &=& \sqrt{\Sigma(x)}  \sum_{n_2, \dots, n_p} \frac{\langle n_2 | \Sigma | n_3\rangle \dots \langle n_p | \Sigma | 1\rangle}{\epsilon_{n_2} \dots \epsilon_{n_p}}  \phi_{n_2}(x_1,\dots,x_d) \nonumber\\ 
&=&  \Xi^{(p-1)}(x_1, \dots, x_d)
\eeq
which confirms that the operator in Eq.~(\ref{app_A_eq2}) is indeed the inverse operator of $\hat{O}$.

It useful to discuss also the case in which the zero mode is present. In this case we need to apply Theorem \ref{theo1zero}, starting from an ansatz
\begin{eqnarray}
\tilde{\Xi}^{(0)}({\bf x}) &=& \sqrt{\Sigma({\bf x})} \phi_{\bf n_0}({\bf x})
\end{eqnarray}
which is then orthogonalized with respect to the zero mode:
\begin{eqnarray}
\Xi^{(0)}({\bf x}) &=& \sqrt{\Sigma({\bf x})} \left[ \phi_{\bf n_0}({\bf x}) -  
\frac{1}{\sqrt{\Omega_d}}\frac{\langle {\bf n_0} |\Sigma | 0 \rangle}{\langle 0 |\Sigma | 0 \rangle } \right]
\end{eqnarray}

After one iteration we have
\beq
\tilde{\Xi}^{(1)}({\bf x}) &=& \sqrt{\Sigma({\bf x})} \sum_{n_1}^\prime \frac{\phi_{{\bf n}_1} ({\bf x}) }{\epsilon_{{\bf n}_1}} \left[ 
\langle {\bf n}_1 |\Sigma | {\bf n}_0 \rangle - \frac{\langle {\bf n}_1 |\Sigma | 0 \rangle \langle {\bf n}_0 |\Sigma | 0 \rangle}{\langle 0 |\Sigma | 0 \rangle}
\right]
\eeq
and
\begin{eqnarray}
\Xi^{(1)}({\bf x}) &=& \sqrt{\Sigma({\bf x})}  \sum_{n_1}^\prime \frac{1}{\epsilon_{{\bf n}_1}}  
\left( \phi_{{\bf n}_1}({\bf x}) - \frac{1}{\sqrt{\Omega_d}}\frac{\langle {\bf n}_1 |\Sigma | 0 \rangle}{\langle 0 |\Sigma | 0 \rangle } \right) \cdot
\left( \langle {\bf n}_1 |\Sigma | {\bf n}_0 \rangle - \frac{\langle {\bf n}_1 |\Sigma | 0 \rangle \langle {\bf n}_0 |\Sigma | 0 \rangle}{\langle 0 |\Sigma | 0 \rangle}
\right)
\end{eqnarray}

After two iterations we have
\beq
\tilde{\Xi}^{(2)}({\bf x}) &=& \sqrt{\Sigma({\bf x})} \sum_{n_1,n_2}^\prime \frac{\phi_{{\bf n}_1} ({\bf x}) }{\epsilon_{{\bf n}_1} \epsilon_{{\bf n}_2}} 
\left( \langle {\bf n}_1 |\Sigma | {\bf n}_2 \rangle - \frac{\langle {\bf n}_1 |\Sigma | 0 \rangle \langle {\bf n}_2 |\Sigma | 0 \rangle}{\langle 0 |\Sigma | 0 \rangle}
\right) \nonumber \\
&\cdot& \left( \langle {\bf n}_2 |\Sigma | {\bf n}_0 \rangle - \frac{\langle {\bf n}_2 |\Sigma | 0 \rangle \langle {\bf n}_0 |\Sigma | 0 \rangle}{\langle 0 |\Sigma | 0 \rangle} \right)
\eeq
and
\beq
\Xi^{(2)}({\bf x}) &=& \sqrt{\Sigma({\bf x})} \sum_{n_1,n_2}^\prime \frac{1}{\epsilon_{{\bf n}_1} \epsilon_{{\bf n}_2}} 
\left( \phi_{{\bf n}_1}({\bf x}) - \frac{1}{\sqrt{\Omega_d}}\frac{\langle {\bf n}_1 |\Sigma | 0 \rangle}{\langle 0 |\Sigma | 0 \rangle } \right) 
\left( \langle {\bf n}_1 |\Sigma | {\bf n}_2 \rangle - \frac{\langle {\bf n}_1 |\Sigma | 0 \rangle \langle {\bf n}_2 |\Sigma | 0 \rangle}{\langle 0 |\Sigma | 0 \rangle}
\right) \nonumber \\
&\cdot& \left( \langle {\bf n}_2 |\Sigma | {\bf n}_0 \rangle - \frac{\langle {\bf n}_2 |\Sigma | 0 \rangle \langle {\bf n}_0 |\Sigma | 0 \rangle}{\langle 0 |\Sigma | 0 \rangle} \right)
\eeq

Similarly, after $p$ iterations, one has
\beq
\tilde{\Xi}^{(p)}({\bf x}) &=& \sqrt{\Sigma({\bf x})} \sum_{n_1,\dots,n_p}^\prime \frac{\phi_{{\bf n}_1} ({\bf x}) }{\epsilon_{{\bf n}_1} \dots \epsilon_{{\bf n}_p}} 
\left( \langle {\bf n}_1 |\Sigma | {\bf n}_2 \rangle - \frac{\langle {\bf n}_1 |\Sigma | 0 \rangle \langle {\bf n}_2 |\Sigma | 0 \rangle}{\langle 0 |\Sigma | 0 \rangle}
\right) \dots \nonumber \\
&\cdot& \left( \langle {\bf n}_p |\Sigma | {\bf n}_0 \rangle - \frac{\langle {\bf n}_p |\Sigma | 0 \rangle \langle {\bf n}_0 |\Sigma | 0 \rangle}{\langle 0 |\Sigma | 0 \rangle} \right)
\eeq
and
\beq
\Xi^{(p)}({\bf x}) &=& \sqrt{\Sigma({\bf x})} \sum_{n_1,\dots,n_p}^\prime \frac{1}{\epsilon_{{\bf n}_1} \dots \epsilon_{{\bf n}_p}} 
\left( \phi_{{\bf n}_1}({\bf x}) - \frac{1}{\sqrt{\Omega_d}}\frac{\langle {\bf n}_1 |\Sigma | 0 \rangle}{\langle 0 |\Sigma | 0 \rangle } \right) 
\left( \langle {\bf n}_1 |\Sigma | {\bf n}_2 \rangle - \frac{\langle {\bf n}_1 |\Sigma | 0 \rangle \langle {\bf n}_2 |\Sigma | 0 \rangle}{\langle 0 |\Sigma | 0 \rangle}
\right) \dots \nonumber \\
&\cdot& \left( \langle {\bf n}_p |\Sigma | {\bf n}_0 \rangle - \frac{\langle {\bf n}_p |\Sigma | 0 \rangle \langle {\bf n}_0 |\Sigma | 0 \rangle}{\langle 0 |\Sigma | 0 \rangle} \right)
\eeq

Let us now apply the operator $\hat{O}$ to $\Xi^{(p)}({\bf x})$, using the properties discussed for the previous case:
\beq
\hat{O} \Xi^{(p)}({\bf x}) &=& \frac{1}{\sqrt{\Sigma({\bf x})} }
\sum_{n_1,\dots,n_p}^\prime \frac{1}{\epsilon_{{\bf n}_2} \dots \epsilon_{{\bf n}_p}} 
\phi_{{\bf n}_1}({\bf x})  
\left( \langle {\bf n}_1 |\Sigma | {\bf n}_2 \rangle - \frac{\langle {\bf n}_1 |\Sigma | 0 \rangle \langle {\bf n}_2 |\Sigma | 0 \rangle}{\langle 0 |\Sigma | 0 \rangle}
\right) \dots \nonumber \\
&\cdot& \left( \langle {\bf n}_p |\Sigma | {\bf n}_0 \rangle - \frac{\langle {\bf n}_p |\Sigma | 0 \rangle \langle {\bf n}_0 |\Sigma | 0 \rangle}{\langle 0 |\Sigma | 0 \rangle} \right)
\eeq

Now, using the completeness of the basis, one has
\beq
\sum_{n_1,\dots,n_p}^\prime \phi_{{\bf n}_1}({\bf x})  
\left( \langle {\bf n}_1 |\Sigma | {\bf n}_2 \rangle - \frac{\langle {\bf n}_1 |\Sigma | 0 \rangle \langle {\bf n}_2 |\Sigma | 0 \rangle}{\langle 0 |\Sigma | 0 \rangle}
\right) = \Sigma({\bf x}) \ \left(  \phi_{{\bf n}_2}({\bf x}) - \frac{1}{\sqrt{\Omega_d}}\frac{\langle {\bf n}_2 |\Sigma | 0 \rangle}{\langle 0 |\Sigma | 0 \rangle }\right)
\eeq
and therefore
\beq
\hat{O} \Xi^{(p)}({\bf x}) &=& \sqrt{\Sigma({\bf x})} 
\sum_{n_2,\dots,n_p}^\prime \frac{1}{\epsilon_{{\bf n}_2} \dots \epsilon_{{\bf n}_p}} 
\left(  \phi_{{\bf n}_2}({\bf x}) - \frac{1}{\sqrt{\Omega_d}}\frac{\langle {\bf n}_2 |\Sigma | 0 \rangle}{\langle 0 |\Sigma | 0 \rangle }\right)
\left( \langle {\bf n}_2 |\Sigma | {\bf n}_3 \rangle - \frac{\langle {\bf n}_2 |\Sigma | 0 \rangle \langle {\bf n}_3 |\Sigma | 0 \rangle}{\langle 0 |\Sigma | 0 \rangle}
\right) \dots \nonumber \\
&\cdot& \left( \langle {\bf n}_p |\Sigma | {\bf n}_0 \rangle - \frac{\langle {\bf n}_p |\Sigma | 0 \rangle \langle {\bf n}_0 |\Sigma | 0 \rangle}{\langle 0 |\Sigma | 0 \rangle} \right) =  \Xi^{(p-1)}({\bf x})
\eeq
which confirms the conclusion in the previous appendix that  operator in Eq.~(\ref{app_A_eq2}) is indeed the inverse operator of $\hat{O}$ for the class of 
functions that we are considering.

\section*{Acknowledgements}
This research was supported by the Sistema Nacional de Investigadores (M\'exico). The plots contained in this paper have been produced using {TikZ} \cite{TikZ}.

\end{document}